\documentclass[doc,a4paper]{apa6}

\usepackage{url,tabularx,array}
\usepackage{amsmath}
\usepackage{amssymb} 
\usepackage{graphicx}
\usepackage{subfig}
\usepackage{color}
\usepackage{bm}
\usepackage{array}
\usepackage{natbib}

\newcommand\blfootnote[1]{%
  \begingroup
  \renewcommand\thefootnote{}\footnote{#1}%
  \addtocounter{footnote}{-1}%
  \endgroup
}

\usepackage{array}
\newcolumntype{L}[1]{>{\raggedright\let\newline\\\arraybackslash\hspace{0pt}}m{#1}}
\newcolumntype{C}[1]{>{\centering\let\newline\\\arraybackslash\hspace{0pt}}m{#1}}
\newcolumntype{R}[1]{>{\raggedleft\let\newline\\\arraybackslash\hspace{0pt}}m{#1}}

\usepackage{amsmath}
\usepackage{amsfonts}

\newcommand\independent{\protect\mathpalette{\protect\independenT}{\perp}}
\def\independenT#1#2{\mathrel{\rlap{$#1#2$}\mkern2mu{#1#2}}}

\usepackage[american]{babel}
\usepackage{csquotes}

\newcommand{\textVerb}[1]{\texttt{\mbox{#1}}}

\title{Network Psychometrics}

\shorttitle{NETWORK PSYCHOMETRICS}

\author{
Sacha Epskamp, 
Gunter K.\ J.\  Maris,
Lourens J.\ Waldorp and Denny Borsboom
}

\affiliation{University of Amsterdam, Department of Psychological Methods}

\abstract{
This chapter provides a general introduction of network modeling in psychometrics. The chapter starts with an introduction to the statistical model formulation of pairwise Markov random fields (PMRF), followed by an introduction of the PMRF suitable for binary data: the \emph{Ising model}. The Ising model is a model used in ferromagnetism to explain phase transitions in a field of particles. Following the description of the Ising model in statistical physics, the chapter continues to show that the Ising model is closely related to models used in psychometrics. The Ising model can be shown to be equivalent to certain kinds of logistic regression models, loglinear models and multi-dimensional item response theory (MIRT) models. The equivalence between the Ising model and the MIRT model puts standard psychometrics in a new light and leads to a strikingly different interpretation of well-known latent variable models. The chapter gives an overview of methods that can be used to estimate the Ising model, and concludes with a discussion on the interpretation of latent variables given the equivalence between the Ising model and MIRT.
}

\begin{document}

\maketitle\blfootnote{Please cite as: Epskamp, S., Maris, G., Waldorp, L.J., and Borsboom, D.\ (in press).
Network Psychometrics. In Irwing, P., Hughes, D., and Booth, T.\ (Eds.),
\textit{Handbook of Psychometrics}. New York: Wiley.}

\raggedbottom

\begin{quote}
In fact, statistical field theory may have even more to offer. It always struck me that there appears to be a close connection between the basic expressions underlying item-response theory and the solutions of elementary lattice fields in statistical physics. For instance, there is almost a one-to-one formal correspondence of the solution of the Ising model (a lattice with nearest neighbor interaction between binary-valued sites; e.g., \citealt{kindermann1980markov}, Chapter 1) and the Rasch model \citep{fischer1974einfuhrung}. \\ \par\raggedleft---\textup{Peter Molenaar} (\citeyear{molenaar2003state}, p.\ 82)
\end{quote}

\section{Introduction}

In recent years, network models have been proposed as an alternative way of looking at psychometric problems \citep{van2006dynamical,cramer2010comorbidity, borsboom2013network}. In these models, psychometric item responses are conceived of as proxies for variables that directly interact with each other. For example, the symptoms of depression (such as loss of energy, sleep problems, and low self esteem) are traditionally thought of as being determined by a common latent variable (depression, or the liability to become depressed; \citealt{aggen2005dsm}). In network models, these symptoms are instead hypothesized to form networks of mutually reinforcing variables (e.g., sleep problems may lead to loss of energy, which may lead to low self esteem, which may cause rumination that in turn may reinforce sleep problems). On the face of it, such network models offer an entirely different conceptualization of why psychometric variables cluster in the way that they do. However, it has also been suggested in the literature that latent variables may somehow correspond to sets of tightly intertwined observables (e.g., see the Appendix of \citealt{van2006dynamical}), and as the above quote shows, \citet{molenaar2003state} already suspected that network models in physics are closely connected to psychometric models with latent variables. 

In the current chapter, we aim to make this connection explicit. As we will show, a particular class of latent variable models (namely, multidimensional Item Response Theory models) yields exactly the same probability distribution over the observed variables as a particular class of network models (namely, Ising models). In the current chapter, we exploit the consequences of this equivalence. We will first introduce the general class of models used in network analysis called Markov Random Fields. Specifically, we will discuss the Markov random field for binary data called the \emph{Ising Model}, which originated from statistical physics but has since been used in many fields of science. We will show how the Ising Model relates to psychometrical practice, with a focus on the equivalence between the Ising Model and multidimensional item response theory. We will demonstrate how the Ising model can be estimated and finally, we will discuss the conceptual implications of this equivalence.

\subsection{Notation}

Throughout this chapter we will denote random variables with capital letters and possible realizations with lower case letters; vectors will be represented with bold-faced letters. For parameters, we will use boldfaced capital letters to indicate matrices instead of vectors whereas for random variables we will use boldfaced capital letters to indicate a random vector. Roman letters will be used to denote observable variables and parameters (such as the number of nodes) and Greek letters will be used to denote unobservable variables and parameters that need to be estimated. 

In this chapter we will mainly model the random vector $\boldsymbol{X}$:
\[
\boldsymbol{X}^\top =  \begin{bmatrix} X_1 & X_2 & \ldots & X_P \end{bmatrix},
\] 
containing $P$ binary variables that take the values $1$ (e.g., correct, true or yes) and $-1$ (e.g., incorrect, false or no). We will denote a realization, or \emph{state}, of $\boldsymbol{X}$ with $\boldsymbol{x}^\top =  \begin{bmatrix} x_1 & x_2 & \ldots & x_p \end{bmatrix}$. Let $N$ be the number of observations and $n( \pmb{x})$ the number of observations that have response pattern $\pmb{x}$. Furthermore, let $i$ denote the subscript of a random variable and $j$ the subscript of a different random variable ($j \not= i$). Thus, $X_i$ is the $i$th random variable and $x_i$ its realization. The superscript $-(\dots)$ will indicate that elements are removed from a vector; for example, $\boldsymbol{X}^{-(i)}$ indicates the random vector $\pmb{X}$ without $X_i$: $\boldsymbol{X}^{-(i)} = \begin{bmatrix} X_1, \ldots, X_{i-1}, X_{i+1}, \ldots. X_P\end{bmatrix}$, and $\boldsymbol{x}^{-(i)}$ indicates its realization. Similarly,  $\boldsymbol{X}^{-(i,j)}$ indicates $\pmb{X}$ without $X_i$ and $X_j$ and $\boldsymbol{x}^{-(i,j)}$ its realization. An overview of all notations used in this chapter can be seen in Appendix~B.

\section{Markov Random Fields}

\begin{figure}
\begin{center}
\includegraphics[width = 0.4\textwidth]{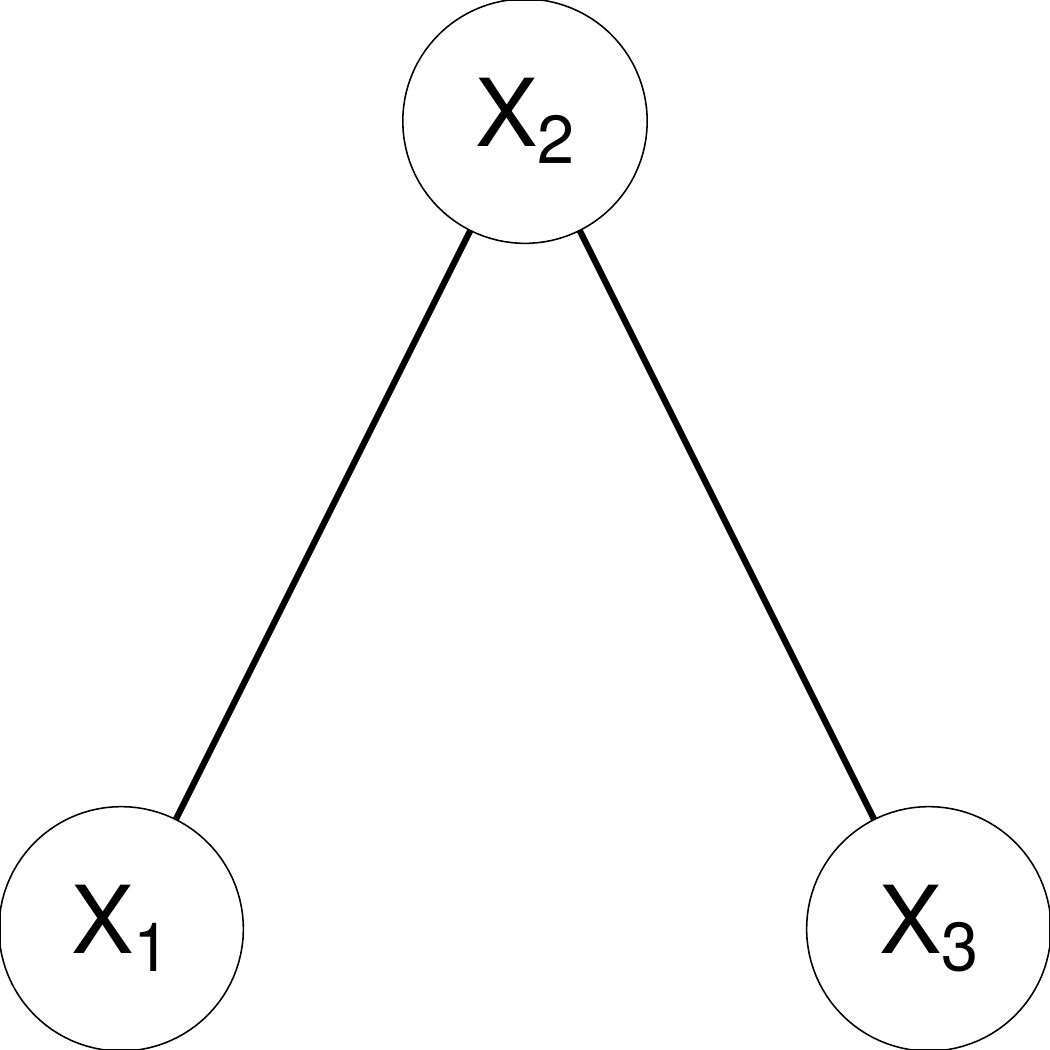}
\end{center}
\caption{Example of a PMRF of three nodes, $X_1$, $X_2$ and $X_3$ , connected by two edges, one between $X_1$ and $X_2$ and one between $X_2$ and $X_3$. }
\label{handbook:fig:mrf}
\end{figure}

A network, also called a graph, can be encoded as a set $G$ consisting of two sets: $V$, which contains the nodes in the network, and $E$, which contains the edges that connect these nodes. For example, the graph in Figure~\ref{handbook:fig:mrf} contains three nodes: $V=\{1,2,3\}$, which are connected by two edges: $E=\{(1,2),(2,3)\}$. We will use this type of network to represent a \emph{pairwise Markov random field} (PMRF; \citealt{lauritzen1996graphical, murphy2012machine}), in which nodes represent observed random variables\footnote{Throughout this chapter, nodes in a network designate variables, hence the terms are used interchangeably.} and edges represent (conditional) association between two nodes. More importantly, the absence of an edge represents the Markov property that two nodes are conditionally independent given all other nodes in the network:
\begin{equation}
\label{handbook:markovproperty}
X_i \independent X_j \mid \boldsymbol{X}^{-(i,j)} = \boldsymbol{x}^{-(i,j)} \iff (i,j) \not\in E
\end{equation}
Thus, a PMRF encodes the independence structure of the system of nodes.  In the case of Figure~\ref{handbook:fig:mrf}, $X_1$ and $X_3$ are independent given that we know $X_2 = x_2$. This could be due to several reasons; there might be a causal path from $X_1$ to $X_3$ or vise versa, $X_2$ might be the common cause of $X_1$ and $X_3$, unobserved variables might cause the dependencies between $X_1$ and $X_2$ and $X_2$ and $X_3$, or the edges in the network might indicate actual pairwise interactions between $X_1$ and $X_2$ and $X_2$ and $X_3$.

Of particular interest to psychometrics are models in which the presence of latent common causes induces associations among the observed variables. If such a common cause model holds, we cannot condition on any observed variable to completely remove the association between two nodes \citep{pearl2000causality}. Thus, if an unobserved variable acts as a common cause to some of the observed variables, we should find a fully connected clique in the PMRF that describes the associations among these nodes. The network in Figure~\ref{handbook:fig:mrf}, for example, cannot represent associations between three nodes that are subject to the influence of a latent common cause; if that were the case, it would be impossible to obtain conditional independence between $X_1$ and $X_3$ by conditioning on $X_2$.

\subsection{Parameterizing Markov Random Fields}

A PMRF can be parameterized as a product of strictly positive potential functions $\phi(x)$ \citep{murphy2012machine}:
\begin{equation}
\label{handbook:MRF}
\Pr \left( \pmb{X} = \pmb{x}\right) = \frac{1}{Z} \prod_{i} \phi_i\left(x_i\right) \prod_{<ij>} \phi_{ij}\left( x_i , x_j\right),
\end{equation}
in which $\prod_{i}$ takes the product over all nodes, $i = 1,2,\ldots,P$,  $\prod_{<ij>}$ takes the product over all distinct pairs of nodes $i$ and $j$ ($j>i$), and $Z$ is a normalizing constant such that the probability function sums to unity over all possible patterns of observations in the sample space:
\[
Z = \sum_{\pmb{x}} \prod_{i} \phi_i\left(x_i\right) \prod_{<ij>} \phi_{ij}\left( x_i , x_j\right).
\]
Here, $\sum_{\pmb{x}}$ takes the sum over all possible realizations of $\pmb{X}$. All $\phi(x)$ functions result in positive real numbers, which encode the \emph{potentials}: the preference for the relevant part of $\pmb{X}$ to be in some state. The $\phi_i(x_i)$ functions encode the node potentials of the network; the preference of node $X_i$ to be in state $x_i$, regardless of the state of the other nodes in the network. Thus, $\phi_i(x_i)$ maps the potential for $X_i$ to take the value $x_i$ regardless of the rest of the network. If $\phi_i(x_i)=0$, for instance, then $X_i$ will never take the value $x_i$, while $\phi_i(x_i)=1$ indicates that there is no preference for $X_i$ to take any particular value and $\phi_i(x_i) = \infty$ indicates that the system always prefers $X_i$ to take the value $x_i$.  The $\phi_{ij}(x_i,x_j)$ functions encode the pairwise potentials of the network; the preference of nodes $X_i$ and $X_j$ to both be in states $x_i$ and $x_j$. As $\phi_{ij}(x_i,x_j)$ grows higher we would expect to observe $X_j=x_j$ whenever $X_i=x_i$. Note that the potential functions are not identified; we can multiply both $\phi_i(x_i)$ or $\phi_{ij}(x_i, x_j)$ with some constant for all possible outcomes of $x_i$, in which case this constant becomes a constant multiplier to \eqref{handbook:MRF} and is cancelled out in the normalizing constant $Z$. A typical identification constraint on the potential functions is to set the marginal geometric means of all outcomes equal to $1$; over all possible outcomes of each argument, the logarithm of each potential function should sum to $0$:
\begin{equation}
\label{handbook:constraints}
\sum_{x_i} \ln \phi_i(x_i) = \sum_{x_i} \ln \phi_{ij}(x_i, x_j) = \sum_{x_j} \ln \phi_{ij}(x_i, x_j) = 0 \quad \forall x_i, x_j
\end{equation}
in which $\sum_{x_i}$ denotes the sum over all possible realizations for $X_i$, and $\sum_{x_j}$ denotes the sum over all possible realizations of $X_j$.

We assume that every node has a potential function $\phi_i(x_i)$ and nodes only have a relevant pairwise potential function $\phi_{ij}(x_i, x_j)$ when they are connected by an edge; thus, two unconnected nodes have a constant pairwise potential function which, due to identification above, is equal to $1$ for all possible realizations of $X_i$ and $X_j$:
\begin{equation}
\label{handbook:noedgepotential}
\phi_{ij}(x_i, x_j) = 1 \quad \forall x_i,x_j \iff (i,j) \not\in E.
\end{equation}

From Equation \eqref{handbook:MRF} it follows that the distribution of $\pmb{X}$ marginalized over $X_k$ and $X_l$, that is, the marginal distribution of $\pmb{X}^{-(k,l)}$ (the random vector $\pmb{X}$ without elements $X_k$ and $X_l$), has the following form:
{\footnotesize
\begin{align}
\nonumber
\Pr \left( \pmb{X}^{-(k,l)} =\pmb{x}^{-(k,l)}\right) &= \sum_{x_k,x_l} \Pr \left(\pmb{X} = \pmb{x}  \right) \\
\label{handbook:marginal}
&= \frac{1}{Z} \prod_{i\not\in \{k,l\}} \phi_i\left(x_i\right) \prod_{<ij\not\in \{k,l\}>} \phi_{ij}\left( x_i , x_j\right) \\
\nonumber &\quad
\sum_{x_k,x_l} \left( \phi_k(x_k) \phi_l(x_l)\phi_{kl}(x_k, x_l) \prod_{i\not\in \{k,l\}} \phi_{ik}(x_i, x_k)  \phi_{il}(x_i, x_l)   \right),
\end{align}
}
in which $ \prod_{i\not\in \{k,l\}}$ takes the product over all nodes except node $k$ and $l$ and $\prod_{<ij\not\in \{k,l\}>} $ takes the product over all unique pairs of nodes that do not involve $k$ and $l$. The expression in \eqref{handbook:marginal} has two important consequences. First, \eqref{handbook:marginal} does not have the form of \eqref{handbook:MRF}; a PMRF is \emph{not} a PMRF under marginalization. Second, dividing \eqref{handbook:MRF} by \eqref{handbook:marginal} an expression can be obtained for the conditional distribution of $\{ X_k, X_l \}$ given that we know $\pmb{X}^{-(k,l)} = \pmb{x}^{-(k,l)}$:
\begin{align}
\nonumber
\Pr \left(X_k, X_l \mid \pmb{X}^{-(k,l)} = \pmb{x}^{-(k,l)} \right) &= \frac{ \Pr \left(\pmb{X} = \pmb{x}  \right)}{ \Pr \left( \pmb{X}^{-(k,l)} =\pmb{x}^{-(k,l)}\right) } \\
&= \frac{\phi_k^*(x_k) \phi_l^*(x_l)\phi_{kl}(x_k, x_l) }{\sum_{x_k,x_l}\phi_k^*(x_k) \phi_l^*(x_l)\phi_{kl}(x_k, x_l) },
\label{handbook:conditional}
\end{align}
in which:
\[
\phi_k^*(x_k) = \phi_k(x_k)\prod_{i\not\in \{k,l\}} \phi_{ik}(x_i, x_k)
\] and: 
\[
\phi_l^*(x_l) = \phi_l(x_l)\prod_{i\not\in \{k,l\}} \phi_{il}(x_i, x_l).
\] Now, \eqref{handbook:conditional} \emph{does} have the same form as \eqref{handbook:MRF}; a PMRF \emph{is} a PMRF under conditioning. Furthermore, if there is no edge between nodes $k$ and $l$, $\phi_{kl}(x_k,x_l) = 1$ according to \eqref{handbook:noedgepotential}, in which case \eqref{handbook:conditional} reduces to a product of two independent functions of $x_k$ and $x_l$ which renders $X_k$ and $X_l$ independent; thus proving the Markov property in \eqref{handbook:markovproperty}.

\subsection{The Ising Model}

The node potential functions $\phi_i(x_i)$ can map a unique potential for every possible realization of $X_i$ and the pairwise potential functions  $\phi_{ij}(x_i, x_j)$ can likewise map unique potentials to every possible pair of outcomes for $X_i$ and $X_j$. When the data are binary, only two realizations are possible for $x_i$, while four realizations are possible for the pair $x_i$ and $x_j$. Under the constraint that the log potential functions should sum to $0$ over all marginals, this means that in the binary case each potential function has one degree of freedom. If we let all $X$'s take the values $1$ and $-1$, there exists a conveniently loglinear model representation for the potential functions:
\begin{align*}
\ln \phi_i(x_i) &= \tau_i x_i \\
\ln \phi_{ij}(x_i, x_j) &= \omega_{ij} x_i x_j.
\end{align*}
The parameters $\tau_i$ and $\omega_{ij}$ are real numbers. In the case that $x_i=1$ and $x_j=1$, it can be seen that these parameters form an identity link with the logarithm of the potential functions:
\begin{align*}
\tau_i &=   \ln \phi_i(1) \\
\omega_{ij} &= \ln \phi_{ij}(1, 1).
\end{align*}
These parameters are centered on $0$ and have intuitive interpretations. The $\tau_i$ parameters can be interpreted as \emph{threshold parameters}. If $\tau_i=0$ the model does not prefer to be in one state or the other, and if $\tau_i$ is higher (lower) the model prefers node $X_i$ to be in state 1 (-1). The $\omega_{ij}$ parameters are the \emph{network parameters} and denote the pairwise interaction between nodes $X_i$ and $X_j$; if $\omega_{ij} = 0$ there is no edge between nodes $X_i$ and $X_j$:
\begin{equation}
\label{handbook:eq:gamma}
\omega_{ij}  
\begin{cases} = 0  &\mbox{if } (i,j) \not\in E \\ 
\in \mathbb{R}  &\mbox{if } (i,j) \in E
\end{cases}.
\end{equation}
The higher (lower) $\omega_{ij}$ becomes, the more nodes $X_i$ and $X_j$ prefer to be in the same (different) state. Implementing these potential functions in \eqref{handbook:MRF} gives the following distribution for $\pmb{X}$:
\begin{align}
\label{handbook:eq:IsingModel}
\Pr\left( \boldsymbol{X} =  \boldsymbol{x} \right) &= \frac{1}{Z} \exp \left( \sum_i \tau_i x_i  + \sum_{<ij>} \omega_{ij} x_i x_j  \right) \\
\nonumber
Z &= \sum_{\boldsymbol{x}} \exp \left( \sum_i \tau_i x_i  + \sum_{<ij>} \omega_{ij} x_i x_j  \right),
\end{align}
which is known as the Ising model \citep{ising1925beitrag}. 

\begin{table}[ht]
\caption{Probability of all states from the network in Figure~\ref{handbook:fig:mrf}.}
\centering
\begin{tabular}{rrrrrr}
  \hline
 $x_1$ & $x_2$ & $x_3$ & Potential & Probability \\ 
  \hline
-1 & -1 & -1 & 3.6693 & 0.3514 \\ 
  1 & -1 & -1 & 1.1052 & 0.1058 \\ 
  -1 & 1 & -1 & 0.4066 & 0.0389 \\ 
  1 & 1 & -1 & 0.9048 & 0.0866 \\ 
  -1 & -1 & 1 & 1.1052 & 0.1058 \\ 
  1 & -1 & 1 & 0.3329 & 0.0319 \\ 
  -1 & 1 & 1 & 0.9048 & 0.0866 \\ 
  1 & 1 & 1 & 2.0138 & 0.1928 \\ 
   \hline
\end{tabular}
\label{handbook:mrftab}
\end{table}

For example, consider the PMRF in Figure~\ref{handbook:fig:mrf}. In this network there are three nodes ($X_1, X_2$ and $X_3$), and two edges (between $X_1$ and $X_2$, and between $X_2$ and $X_3$). Suppose these three nodes are binary, and take the values $1$ and $-1$. We can then model this PMRF as an Ising model with 3 threshold parameters, $\tau_1$, $\tau_2$ and $\tau_3$ and two network parameters, $\omega_{12}$ and $\omega_{23}$. Suppose we set all threshold parameters to $\tau_1 = \tau_2 = \tau_3 = -0.1$, which indicates that all nodes have a general preference to be in the state $-1$. Furthermore we can set the two network parameters to $\omega_{12} = \omega_{23} = 0.5$. Thus, $X_1$ and $X_2$ prefer to be in the same state, and $X_2$ and $X_3$ prefer to be in the same state as well. Due to these interactions, $X_1$ and $X_3$ become associated; these nodes also prefer to be in the same state, even though they are independent once we condition on $X_2$. We can then compute the non-normalized potentials  $\exp \left( \sum_i \tau_i x_i  + \sum_{<ij>} \omega_{ij} x_i x_j  \right)$ for all possible outcomes of $\pmb{X}$ and finally divide that value by the sum over all non-normalized potentials to compute the probabilities of each possible outcome. For instance, for the state $X_1 = -1, X_2 = 1$ and $X_3 = -1$, we can compute the potential as $\exp \left( -0.1 + 0.1 + -0.1 + -0.5 + -0.5 \right) \approx 0.332$. Computing all these potentials and summing them leads to the normalizing constant of $Z \approx 10.443$, which can then be used to compute the probabilities of each state. These values can be seen in Table \ref{handbook:mrftab}. Not surprisingly, the probability $P(X_1 = -1, X_2 = -1, X_3 = -1)$ is the highest probable state in Table \ref{handbook:mrftab}, due to the threshold parameters being all negative. Furthermore, the probability $P(X_1 = 1, X_2 = 1, X_3 = 1)$ is the second highest probability in Table \ref{handbook:mrftab}; if one node is put into state $1$ then all nodes prefer to be in that state due to the network structure.

 \begin{figure}
 \centering
    \subfloat[\label{1}]{%
      \includegraphics[width=0.15\textwidth,page=1]{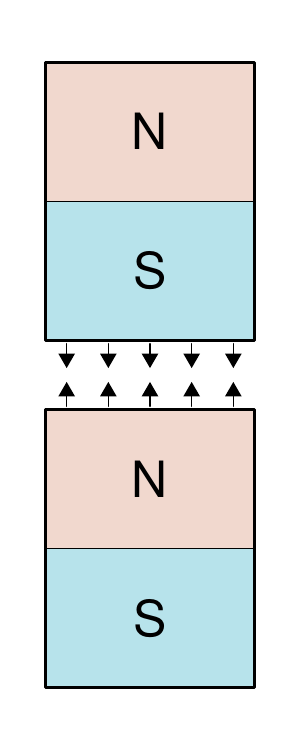}
    }
    \subfloat[\label{2}]{%
      \includegraphics[width=0.15\textwidth,page=2]{fig2}
    } 
    \subfloat[\label{3}]{%
      \includegraphics[width=0.15\textwidth,page=3]{fig2}
    } 
    \subfloat[\label{4}]{%
      \includegraphics[width=0.15\textwidth,page=4]{fig2}
    } 
    \caption{Example of the effect of holding two magnets with a north and south pole close to each other. The arrows indicate the direction the magnets want to move; the same poles, as in (b) and (c), repulse each other and opposite poles, as in (a) and (d), attract each other. }
    \label{handbook:fig:magnets}
  \end{figure}

The Ising model was introduced in statistical physics, to explain the phenomenon of magnetism. To this end, the model was originally defined on a field of particles connected on a lattice. We will give a short introduction on this application in physics because it exemplifies an important aspect of the Ising model; namely, that the interactions between nodes can lead to synchronized behavior of the system as a whole (e.g., spontaneous magnetization). To explain how this works, note that a magnet, such as a common household magnet or the arrow in a compass, has two poles: a north pole and a south pole. Figure~\ref{handbook:fig:magnets} shows the effect of pushing two such magnets together; the north pole of one magnet attracts to the south pole of another magnet and vise versa, and the same poles on both magnets repulse each other. This is due to the generally tendency of magnets to align, called \emph{ferromagnetism}. Exactly the same process causes the arrow of a compass to align with the magnetic field of the Earth itself, causing it to point north. Any material that is ferromagnetic, such as a plate of iron, consists of particles that behave in the same way as magnets; they have a north and south pole and lie in some direction. Suppose the particles can only lie in two directions: the north pole can be up or the south pole can be up. Figure~\ref{handbook:iron} shows a simple 2-dimensional representation of a possible state for a field of $4\times 4$ particles. We can encode each particle as a random variable, $X_i$, which can take the values $-1$ (south pole is up) and $1$ (north pole is up). Furthermore we can assume that the probability of $X_i$ being in state $x_i$ only depends on the direct neighbors (north, south east and west) of particle $i$. With this assumption in place, the system in Figure~\ref{handbook:iron} can be represented as a PMRF on a lattice, as represented in Figure~\ref{handbook:lattice}.

 \begin{figure}
 \centering
    \subfloat[\label{iron}]{%
      \includegraphics[width=0.4\textwidth,page=1]{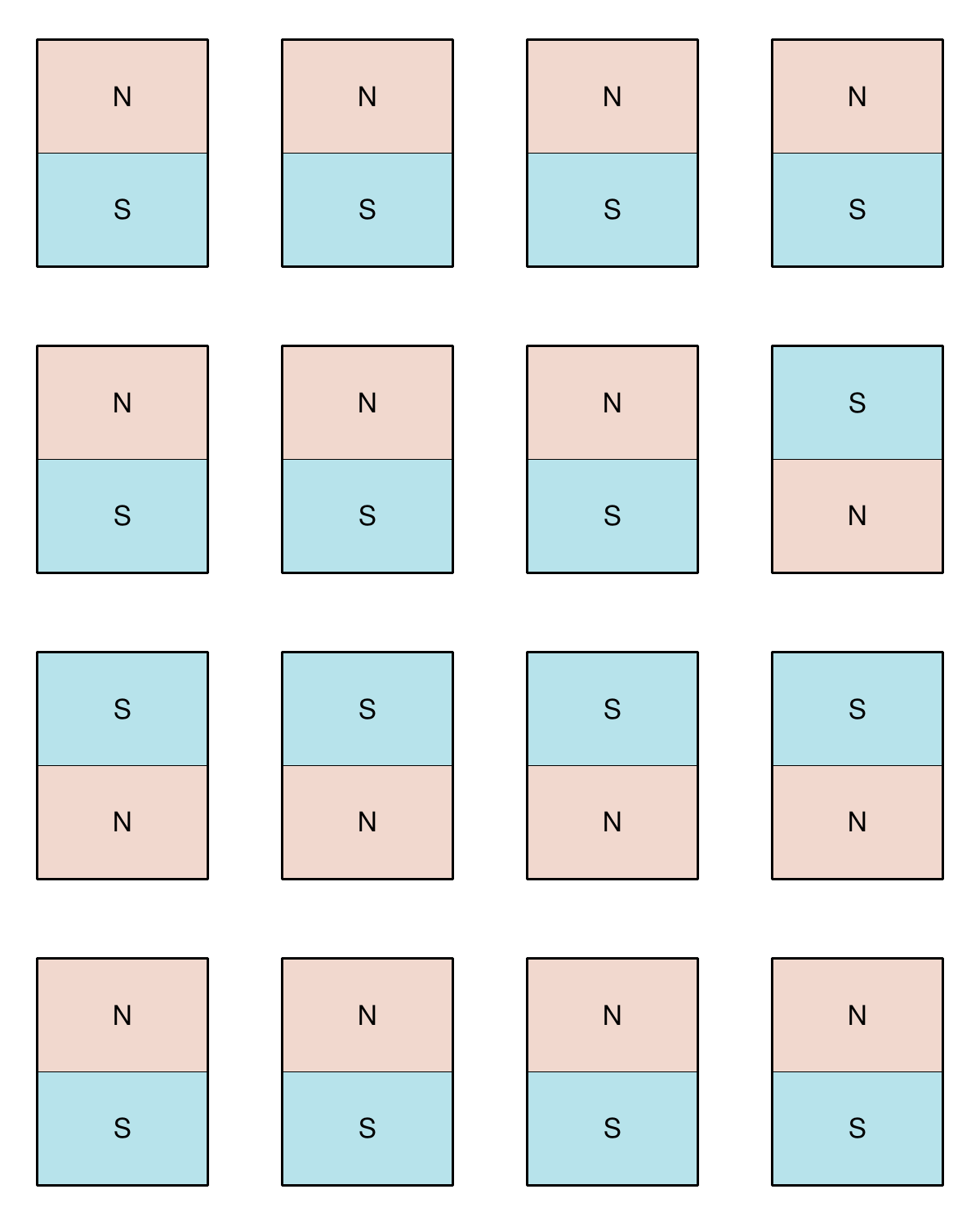}
    }
    \hfill
    \subfloat[\label{lattice}]{%
      \includegraphics[width=0.4\textwidth,page=2]{fig3}
    } 
    \caption{A field of particles (a) can be repressented by a network shaped as a lattice as in (b). $+1$ indicates that the north pole is alligned upwards and $-1$ indicates that the south pole is aligned upwards. The lattice in (b) adheres to a PMRF in that the probability of a particle (node) being in some state is only dependent on the state of its direct neighbors.}
    \label{handbook:fig:latices}
  \end{figure}

A certain amount of energy is required for a system of particles to be in some state, such as in Figure~\ref{handbook:fig:magnets}. For example, in Figure~\ref{handbook:lattice} the node $X_7$ is in the state $-1$ (south pole up).  Its neighbors $X_3$ and $X_{11}$ are both in the same state and thus aligned, which reduces stress on the system and thus reduces the energy function. The other neighbors of $X_7$, $X_6$ and $X_8$, are in the opposite state of $X_7$, and thus are not aligned, which increasing the stress on the system. The total energy configuration can be summarized in the \emph{Hamiltonian} function:
\[
H( \boldsymbol{x} ) = - \sum_i  \tau_i x_i - \sum_{<i,j>} \omega_{ij} x_i x_j,
\]
which is used in the Gibbs distribution \citep{murphy2012machine} to model the probability of $\pmb{X}$ being in some state $\pmb{x}$:
\begin{equation}
\label{handbook:eq:IsingPhysics}
\Pr\left(  \boldsymbol{X} = \boldsymbol{x} \right) = \frac{\exp \left( - \beta H( \boldsymbol{x} ) \right)}{Z}.
\end{equation}
The parameter $\beta$ indicates the inverse temperature of the system, which is not identifiable since we can multiply $\beta$ with some constant and divide all $\tau$ and $\omega$ parameters with that same constant to obtain the same probability. Thus, it can arbitrarily be set to $\beta=1$. Furthermore, the minus signs in the Gibbs distribution and Hamiltonian cancel out, leading to the Ising model as expressed in \eqref{handbook:eq:IsingModel}.

The threshold parameters $\tau_i$ indicate the natural deposition for particle $i$ to point up or down, which could be due to the influence of an external magnetic field not part of the system of nodes in $\pmb{X}$. For example, suppose we model a single compass, there is only one node thus the Hamiltonian reduces to $-\tau x$. Let $X=1$ indicate the compass points north and $X= -1$ indicate the compass points south. Then, $\tau$ should be positive as the compass has a natural tendency to point north due to the presence of the Earth's magnetic field. As such, the $\tau$ parameters are also called external fields. The network parameters $\omega_{ij}$ indicate the interaction between two particles. Its sign indicates if particles $i$ and $j$ tend to be in the same state (positive; ferromagnetic) or in different states (negative; anti-ferromagnetic). The absolute value, $|\omega_{ij}|$, indicates the strength of interaction. For any two non-neighboring particles $\omega_{ij}$ will be $0$ and for neighboring particles the stronger $\omega_{ij}$ the stronger the interaction between the two. Because the closer magnets, and thus particles, are moved together the stronger the magnetic force, we can interpret $|\omega_{ij}|$ as a measure for \emph{closeness} between two nodes.

While the inverse temperature $\beta$ is not identifiable in the sense of parameter estimation, it is an important element in the Ising model; in physics the temperature can be manipulated whereas the ferromagnetic strength or distance between particles cannot. The inverse temperature plays a crucial part in the \emph{entropy} of \eqref{handbook:eq:IsingPhysics} \citep{wainwright2008graphical}:
\begin{align}
\nonumber
\mathrm{Entropy}\left( \pmb{X} \right) &= \mathbb{E}\left[- \ln \Pr\left(  \boldsymbol{X} = \boldsymbol{x} \right) \right] \\
\label{handbook:tempentro}
&= -\beta \mathbb{E}\left[  - \ln  \frac{\exp \left( - H( \boldsymbol{x} ) \right)}{Z^*} \right],
\end{align}
in which $Z^*$ is the rescaled normalizing constant without inverse temperature $\beta$. The expectation $\mathbb{E}\left[  - \ln  \frac{\exp \left( - H( \boldsymbol{x} ) \right)}{Z^*} \right]$ can be recognized as the entropy of the Ising model as defined in \eqref{handbook:eq:IsingModel}. Thus, the inverse temperature $\beta$ directly scales the entropy of the Ising model. As $\beta$ shrinks to $0$, the system is ``heated up'' and all states become equally likely, causing a high level of entropy. If $\beta$ is subsequently increased, then the probability function becomes concentrated on a smaller number of states, and the entropy shrinks to eventually only allow the state in which all particles are aligned. The possibility that all particles become aligned is called \emph{spontaneous magnetization} \citep{lin1992spontaneous, kac1966mathematical}; when all particles are aligned (all $X$ are either $1$ or $-1$) the entire field of particles becomes magnetized, which is how iron can be turned into a permanent magnet. We take this behavior as a particular important aspect of the Ising model; behavior on microscopic level (interactions between neighboring particles) can cause noticeable behavior on macroscopic level (the creation of a permanent magnet).

In our view, psychological variables may behave in the same way. For example, interactions between components of a system (e.g., symptoms of depression) can cause synchronized effects of the system as a whole (e.g., depression as a disorder). Do note that, in setting up such analogies, we need to interpret the concepts of closeness and neighborhood less literally than in the physical sense. Concepts such as ``sleep deprivation'' and ``fatigue'' can be said to be close to each other, in that they mutually influence each other; sleep deprivation can lead to fatigue and in turn fatigue can lead to a disrupted sleeping rhythm. The neighborhood of these symptoms can then be defined as the symptoms that frequently co-occur with sleep deprivation and fatigue, which can be seen in a network as a cluster of connected nodes. As in the Ising model, the state of these nodes will tend to be the same if the connections between these nodes are positive. This leads to the interpretation that a latent trait, such as depression, can be seen as a cluster of connected nodes \citep{borsboom2011small}. In the next section, we will prove that there is a clear relationship between network modeling and latent variable modeling; indeed, clusters in a network can cause data to behave as if they were generated by a latent variable model.

\section{The Ising Model in Psychometrics}

In this section, we show that the Ising model is equivalent or closely related to prominent modeling techniques in psychometrics. We will first discuss the relationship between the Ising model and loglinear analysis and logistic regressions, next show that the Ising model can be equivalent to Item Response Theory (IRT) models that dominate psychometrics. In addition, we highlight relevant earlier work on the relationship between IRT and the Ising model.

To begin, we can gain further insight in the Ising model by looking at the conditional distribution of $X_i$ given that we know the value of the remaining nodes: $\boldsymbol{X}^{(-i)} = \boldsymbol{x}^{(-i)}$:
\begin{align}
\nonumber
\Pr\left( X_i \mid   \boldsymbol{X}^{(-i)} = \boldsymbol{x}^{(-i)} \right) &= \frac{ \Pr\left(   \boldsymbol{X} =  \boldsymbol{x} \right)  }{  \Pr\left(   \boldsymbol{X}^{(-i)} = \boldsymbol{x}^{(-i)} \right) } \\
\nonumber
&= \frac{ \Pr\left(   \boldsymbol{X} =  \boldsymbol{x} \right)  }{\sum_{x_i}    \Pr\left( X_i = x_i,   \boldsymbol{X}^{(-i)} = \boldsymbol{x}^{(-i)} \right)  } \\
\label{handbook:eq:conditional}
&= \frac{ \exp \left( x_i \left(\tau_i  +  \sum_{j} \omega_{ij} x_j  \right) \right) }{ \sum_{x_i} \exp \left( x_i \left(\tau_k  +  \sum_{j} \omega_{ij} x_j  \right) \right)} ,
\end{align}
in which $\sum_{x_i}$ takes the sum over both possible outcomes of $x_i$. We can recognize this expression as a \emph{logistic regression} model \citep{agresti2014categorical}. Thus, the Ising model can be seen as the joint distribution of response and predictor variables, where each variable is predicted by all other variables in the network. The Ising model therefore forms a predictive network in which the neighbors of each node, the set of connected nodes, represent the variables that predict the outcome of the node of interest. 

Note that the definition of Markov random fields in \eqref{handbook:MRF} can be extended to include higher order interaction terms:
\[
\Pr \left( \pmb{X} = \pmb{x}\right) = \frac{1}{Z} \prod_{i} \phi_i\left(x_i\right) \prod_{<ij>} \phi_{ij}\left( x_i , x_j\right)  \prod_{<ijk>} \phi_{ijk}\left( x_i , x_j, x_k \right)  \cdots,
\]
all the way up to the $P$-th order interaction term, in which case the model becomes saturated. Specifying $\nu_{\dots}(\dots) = \ln \phi_{\dots}(\dots)$ for all potential functions, we obtain a log-linear model:
{\footnotesize
\[
\Pr \left( \pmb{X} = \pmb{x}\right) = \frac{1}{Z} \exp \left(  \sum_{i} \nu_i\left(x_i\right) + \sum_{<ij>} \nu_{ij}\left( x_i , x_j\right) + \sum_{<ijk>} \nu_{ijk}\left( x_i , x_j, x_k \right)  \cdots \right).
\]
}

Let $n(\pmb{x})$ be the number of respondents with response pattern $\pmb{x}$ from a sample of $N$ respondents. Then, we may model the expected frequency $n(\pmb{x})$ as follows:
{\footnotesize
\begin{align}
\nonumber
\mathbb{E}\left[ n(\pmb{x}) \right] &= N \Pr \left( \pmb{X} = \pmb{x}\right) \\
\label{handbook:loglin}
&=  \exp \left(\nu +  \sum_{i} \nu_i\left(x_i\right) + \sum_{<ij>} \nu_{ij}\left( x_i , x_j\right) + \sum_{<ijk>} \nu_{ijk}\left( x_i , x_j, x_k \right)  \cdots \right),
\end{align}
}
in which $\nu =  \ln N - \ln Z$. The model in \eqref{handbook:loglin} has extensively been used in loglinear analysis \citep{agresti2014categorical, wickens2014multiway}\footnote{both Agresti and Wickens used $\lambda$ rather than $\nu$ to denote the log potentials, which we changed in this chapter to avoid confusion with eigenvalues and the LASSO tuning parameter.}. In loglinear analysis, the same constrains are typically used as in \eqref{handbook:constraints}; all $\nu$ functions should sum to $0$ over all margins. Thus, if at most second-order interaction terms are included in the loglinear model, it is equivalent to the Ising model and can be represented exactly as in \eqref{handbook:eq:IsingModel}.  The Ising model, when represented as a loglinear model with at most second-order interactions, has been used in various ways. \citet{agresti2014categorical} and \citet{wickens2014multiway} call the model the \emph{homogeneous association} model. Because it does not include three-way or higher order interactions, the association between $X_i$ and $X_j$---the odds-ratio---is constant for any configuration of $\pmb{X}^{-(i,j)}$. Also,  \citeauthor{cox1972analysis} (\citeyear{cox1972analysis};  \citealt{cox1994note}) used the same model, but termed it the \emph{quadratic exponential binary distribution}, which has since often been used in biometrics and statistics (e.g., \citealt{fitzmaurice1993regression, zhao1990correlated}). Interestingly, none of these authors mention the Ising model.

\subsection{The Relation Between the Ising Model and Item Response Theory}

In this section we will show that the Ising model is a closely related modeling framework of Item Response Theory (IRT), which is of central importance to psychometrics. In fact, we will show that the Ising model is equivalent to a special case of the multivariate 2-parameter logistic model (MIRT). However, instead of being hypothesized common causes of the item responses, in our representation the latent variables in the model are \emph{generated} by cliques in the network. 

In IRT, the responses on a set of binary variables $\pmb{X}$ are assumed to be determined by an set of $M$ ($M \leq P$) latent variables $\pmb{\Theta}$:
\[
\pmb{\Theta}^\top = \begin{bmatrix} \Theta_1 & \Theta_2 & \ldots & \Theta_M\end{bmatrix}.
\] 
These latent variables are often denoted as \emph{abilities}, which betrays the roots of the model in educational testing. In IRT, the probability of obtaining a realization $x_i$ on the variable $X_i$---often called \emph{items}---is modeled through item response functions, which model the probability of obtaining one of the two possible responses (typically, scored $1$ for correct responses and $0$ for incorrect responses) as a function of $\pmb{\theta}$. For instance, in the \citet{rasch1960studies} model, also called the one parameter logistic model (1PL), only one latent trait is assumed ($M=1$ and $\pmb{\Theta} = \Theta$) and the conditional probability of a response given the latent trait takes the form of a simple logistic function:
\[
\Pr(X_i = x_i \mid \Theta = \theta)_{\mathrm{1PL}} = \frac{\exp\left(x_i \alpha \left( \theta - \delta_i \right) \right)}{\sum_{x_i} \exp\left(x_i \alpha \left( \theta - \delta_i \right) \right)},
\]
in which $\delta_i$ acts as a \emph{difficulty parameter} and $\alpha$ is a common \emph{discrimination} parameter for all items. A typical generalization of the 1PL is the \citet{birnbaum1968some} model, often called the two-parameter logistic model (2PL), in which the discrimination is allowed to vary between items:
\[
\Pr(X_i = x_i \mid \Theta = \theta)_{\mathrm{2PL}} = \frac{\exp\left(x_i \alpha_i \left( \theta - \delta_i \right) \right)}{\sum_{x_i} \exp\left(x_i \alpha_i \left( \theta - \delta_i \right) \right)}.
\]
The 2PL reduces to the 1PL if all discrimination parameters are equal: $\alpha_1 = \alpha_2 = \ldots = \alpha$. Generalizing the 2PL model to more than 1 latent variable ($M>1$) leads to the 2PL multidimensional IRT model (MIRT; \citealt{reckase2009multidimensional}):
\begin{equation}
\label{handbook:MIRTprob}
\Pr(X_i = x_i \mid \pmb{\Theta} = \pmb{\theta})_{\mathrm{MIRT}} = \frac{\exp\left(x_i \left( \pmb{\alpha}_i^{\top}  \pmb{\theta} - \delta_i \right) \right)}{\sum_{x_i} \exp\left(x_i \left( \pmb{\alpha}_i^{\top}  \pmb{\theta} - \delta_i \right) \right)},
\end{equation}
in which $\pmb{\theta}$ is a vector of length $M$ that contains the realization of $\pmb{\Theta}$, while $\pmb{\alpha}_i$ is a vector of length $M$ that contains the discrimination of item $i$ on every latent trait in the multidimensional space. The MIRT model reduces to the 2PL model if $\pmb{\alpha}_i$ equals zero in all but one of its elements. 

Because IRT assumes local independence---the items are independent of each other after conditioning on the latent traits---the joint conditional probability of $\pmb{X} = \pmb{x}$ can be written as product of the conditional probabilities of each item:
\begin{equation}
\label{handbook:jointconditional}
\Pr( \pmb{X} = \pmb{x} \mid \pmb{\Theta} = \pmb{\theta} ) = \prod_i \Pr( X_i = x_i \mid \pmb{\Theta} = \pmb{\theta} ).
\end{equation}
The marginal probability, and thus the likelihood, of the 2PL MIRT model can be obtained by integrating over distribution $f(\pmb{\theta})$ of $\pmb{\Theta}$:
\begin{equation}
\label{handbook:MIRTlikelihood}
 \Pr(\pmb{X} = \pmb{x}) =  \int_{-\infty}^{\infty}  f(\pmb{\theta}) \Pr( \pmb{X} = \pmb{x} \mid \pmb{\Theta} = \pmb{\theta} )  \, \mathrm{d} \pmb{\theta},
\end{equation}
in which the integral is over all $M$ latent variables. For typical distributions of $\pmb{\Theta}$, such as a multivariate Gaussian distribution, this likelihood does not have a closed form solution. Furthermore, as $M$ grows it becomes hard to numerically approximate \eqref{handbook:MIRTlikelihood}. However, if the distribution of $\pmb{\Theta}$ is chosen such that it is conditionally Gaussian---the posterior distribution of $\pmb{\Theta}$ given that we observed $\pmb{X}= \pmb{x}$ takes a Gaussian form---we \emph{can} obtain a closed form solution for \eqref{handbook:MIRTlikelihood}. Furthermore, this closed form solution is, in fact, the Ising model as presented in \eqref{handbook:eq:IsingModel}. 

As also shown by \citet{marsman2015bayesian} and in more detail in Appendix~A of this chapter, after reparameterizing $\tau_i = -\delta_i$ and $-2\sqrt{\lambda_j/2}q_{ij} = \alpha_{ij}$, in which $q_{ij}$ is the $i$th element of the $j$th eigenvector of $\pmb{\Omega}$ (with an arbitrary diagonal chosen such that  $\pmb{\Omega}$ is positive definite), the Ising model is equivalent to a MIRT model in which the posterior distribution of the latent traits is equal to the product of univariate normal distributions with equal variance:
\[
\Theta_j \mid \boldsymbol{X} = \boldsymbol{x} \sim N\left( \pm \frac{1}{2} \sum_i a_{ij} x_i ,   \sqrt{\frac{1}{2}} \right).
\]
The mean of these univariate posterior distributions for $\Theta_j$ is equal to the weighted sumscore $\pm \frac{1}{2} \sum_i a_{ij} x_i$. Finally, since
\[
 f(\pmb{\theta}) = \sum_{\pmb{x}} f( \pmb{\theta} \mid \pmb{X} = \pmb{x})\Pr(\pmb{X} = \pmb{x}),
\]
we can see that the marginal distribution of $\pmb{\Theta}$ in \eqref{handbook:MIRTlikelihood} is a \emph{mixture of multivariate Gaussian distributions with homogenous variance--covariance}, with the mixing probability equal to the marginal probability of observing each response pattern.

Whenever $\alpha_{ij} = 0$ for all $i$ and some dimension $j$---i.e., none of the items discriminate on the latent trait---we can see that the marginal distribution of $\Theta_j$ becomes a Gaussian distribution with mean $0$ and standard-deviation $\sqrt{1/2}$. This corresponds to complete randomness; all states are equally probable given the latent trait. When discrimination parameters diverge from $0$, the probability function becomes concentrated on particular response patterns. For example, in case $X_1$ designates the response variable for a very easy item, while $X_2$ is the response variable for a very hard item, the state in which the first item is answered correctly and the second incorrectly becomes less likely. This corresponds to a decrease in entropy and, as can be seen in \eqref{handbook:tempentro}, is related to the \emph{temperature} of the system. The lower the temperature, the more the system prefers to be in states in which all items are answered correctly or incorrectly. When this happens, the distribution of $\Theta_j$ diverges from a Gaussian distribution and becomes a bi-modal distribution with two peaks, centered on the weighted sumscores that correspond to situations in which all items are answered correctly or incorrectly. If the entropy is relatively high, $f(\Theta_j)$ can be well approximated by a Gaussian distribution, whereas if the entropy is (extremely) low a mixture of two Gaussian distributions best approximates $f(\Theta_j)$.

\begin{figure}
 \centering
    \subfloat[\label{dista}]{%
      \includegraphics[width=0.33\textwidth,page=1]{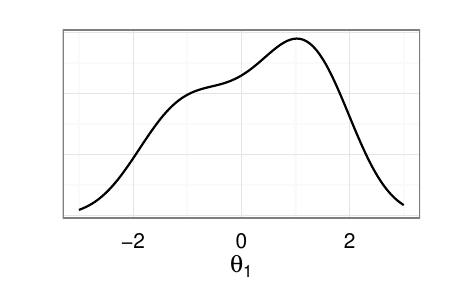}
    }
    \subfloat[\label{distb}]{%
      \includegraphics[width=0.33\textwidth,page=2]{fig4}
    } 
    \subfloat[\label{distc}]{%
      \includegraphics[width=0.33\textwidth,page=3]{fig4}
    } 
    \caption{The distributions of the three latent traits in the equivalent MIRT model to the Ising model from Figure \ref{fig:mrf}}
    \label{handbook:fig:distributions}
  \end{figure}

For example, consider again the network structure of Figure~\ref{handbook:fig:mrf}. When we parameterized all threshold functions $\tau_1 = \tau_2 = \tau_3 = -0.1$ and all network parameters $\omega_{12} = \omega_{23} = 0.5$ we obtained the probability distribution as specified in Table \ref{handbook:mrftab}. We can form the matrix $\pmb{\Omega}$ first with zeroes on the diagonal:
\[
\begin{bmatrix}
0 & 0.5 & 0 \\
0.5 & 0 & 0.5 \\
0 & 0.5 & 0
\end{bmatrix},
\]
which is not positive semi-definite. Subtracting the lowest eigenvalue, $-0.707$, from the diagonal gives us a positive semi-definite  $\pmb{\Omega}$  matrix:
\[
\pmb{\Omega} = \begin{bmatrix}
0.707 & 0.5 & 0 \\
0.5 & 0.707 & 0.5 \\
0 & 0.5 & 0.707
\end{bmatrix}.
\]
It's eigenvalue decomposition is as follows:
\begin{align*}
\pmb{Q} &= \begin{bmatrix}
0.500 & 0.707 & 0.500 \\ 
0.707 & 0.000 & -0.707 \\ 
0.500 & -0.707 & 0.500 \\ 
\end{bmatrix} \\
\pmb{\lambda} &= \begin{bmatrix} 1.414 &  0.707 & 0.000 \end{bmatrix}.
\end{align*}
Using the transformations  $\tau_i = -\delta_i$ and $-2\sqrt{\lambda_j/2}q_{ij} = \alpha_{ij}$ (arbitrarily using the negative root) defined above we can then form the equivalent MIRT model with discrimination parameters $\pmb{A}$ and difficulty parameters $\pmb{\delta}$:
\begin{align*}
\pmb{\delta} &= \begin{bmatrix} 0.1 & 0.1 & 0.1 \end{bmatrix} \\
\pmb{A} &= 
\begin{bmatrix} 
 0.841 & 0.841 & 0 \\ 
 1.189 & 0 & 0 \\ 
 0.841 & -0.841 & 0 \\ \end{bmatrix} .
\end{align*}
Thus, the model in Figure~\ref{handbook:fig:mrf} is equivalent to a model with two latent traits: one defining the general coherence between all three nodes and one defining the contrast between the first and the third node. The distributions of all three latent traits can be seen in Figure~\ref{handbook:fig:distributions}. In Table \ref{handbook:mrftab}, we see that the probability is the highest for the two states in which all three nodes take the same value. This is reflected in the distribution of the first latent trait in \ref{handbook:dista}: because all discrimination parameters relating to this trait are positive, the weighted sumscores of $X_1 = X_2 = X_3 = -1$ and $X_1 = X_2 = X_3 = 1$ are dominant and cause a small bimodality in the distribution. For the second trait, \ref{handbook:distb} shows an approximately normal distribution, because this trait acts as a contrast and cancels out the preference for all variables to be in the same state. Finally, the third latent trait is nonexistent, since all of its discrimination parameters equal $0$; \ref{handbook:distc} simply shows a Gaussian distribution with standard deviation $\sqrt{\frac{1}{2}}$.

This proof serves to demonstrate that the Ising model is equivalent to a MIRT model with a posterior Gaussian distribution on the latent traits; the discrimination parameter column vector $\pmb{\alpha_j}$---the item discrimination parameters on the $j$th dimension---is directly related to the $j$th eigenvector of the Ising model graph structure $\pmb{\Omega}$, scaled by its $j$th eigenvector. Thus, the latent dimensions are orthogonal, and the rank of $\pmb{\Omega}$ directly corresponds to the number of latent dimensions. In the case of a Rasch model, the rank of $\pmb{\Omega}$ should be $1$ and all $\omega_{ij}$ should have exactly the same value, corresponding to the common discrimination parameter; for the uni-dimensional Birnbaum model the rank of $\pmb{\Omega}$ still is $1$ but now the $\omega_{ij}$ parameters can vary between items, corresponding to differences in item discrimination.

The use of a posterior Gaussian distribution to obtain a closed form solution for \eqref{handbook:MIRTlikelihood} is itself not new in the psychometric literature, although it has not previously been linked to the Ising model and the literature related to it. \citet{olkin1961multivariate} already proposed to model binary variables jointly with conditional Gaussian distributed continuous variables. Furthermore,  \citet{holland1990dutch} used the ``Dutch identity'' to show that a representation equivalent to an Ising model could be used to characterize the marginal distribution of an extended Rasch model \citep{cressie1983characterizing}. Based on these results, Anderson and colleagues proposed an IRT modeling framework using log-multiplicative association models and assuming conditional Gaussian latents \citep{anderson2000log, anderson2007log}; this approach has been implemented in the R package ``plRasch'' \citep{anderson2007estimation, plRasch}. 

With our proof we furthermore show that the clique factorization of the network structure \emph{generated} a latent trait with a functional distribution through a mathematical trick.  Thus, the network perspective and common cause perspectives could be interpreted as two different explanations of the same phenomena: cliques of correlated observed variables. In the next section, we show how the Ising model can be estimated.

\section{Estimating the Ising Model}

We can use \eqref{handbook:eq:IsingModel} to obtain the log-likelihood function of a realization $\pmb{x}$:
\begin{align}
\label{handbook:loglikelihood}
\mathcal{L}\left(\pmb{\tau}, \pmb{\Omega} ; \pmb{x} \right)  &= \ln \Pr\left(  \pmb{X} = \pmb{x} \right)=
 \sum_i \tau_i x_{i}  + \sum_{<ij>} \omega_{ij} x_{i} x_{j}   - \ln Z  .
\end{align}

Note that the constant $Z$ is only constant with regard to $\pmb{x}$ (as it sums over all possible realizations) and is \emph{not} a constant with regard to the $\tau$ and $\omega$ parameters; $Z$ is often called the \emph{partition function} because it is a function of the parameters. Thus, while when sampling from the Ising distribution $Z$ does not need to be evaluated, but it \emph{does} need to be evaluated when maximizing the likelihood function. Estimating the Ising model is notoriously hard because the partition function $Z$ is often not tractable to compute \citep{kolaczyk2009statistical}. As can be seen in \eqref{handbook:eq:IsingModel}, $Z$ requires a sum over all possible configurations of $\pmb{x}$; computing $Z$ requires summing over $2^k$ terms, which quickly becomes intractably large as $k$ grows. Thus, maximum likelihood estimation of the Ising model is only possible for trivially small data sets (e.g., $k < 10$). For larger data sets, different techniques are required to estimate the parameters of the Ising model. Markov samplers can be used to estimate the Ising model by either approximating $Z$ \citep{sebastiani2002bayesian,green2002hidden,dryden2003bayesian} or circumventing $Z$ entirely via sampling auxiliary variables  \citep{moller2006efficient, murray2007advances, murray2012mcmc}. Such sampling algorithms can however still be computationally costly.

Because the Ising model is equivalent to the homogeneous association model in log-linear analysis \citep{agresti2014categorical}, the methods used in log-linear analysis can also be used to estimate the Ising model. For example,  the iterative proportional fitting algorithm  \citep{haberman1972algorithm}, which is implemented in the \textVerb{loglin} function in the statistical programming language \emph{R} \citep{R}, can be used to estimate the parameters of the Ising model. Furthermore, log-linear analysis can be used for model selection in the Ising model by setting certain parameters to zero. Alternatively, while the full likelihood in \eqref{handbook:eq:IsingModel} is hard to compute, the conditional likelihood for each node in \eqref{handbook:eq:conditional} is very easy and does not include an intractable normalizing constant; the conditional likelihood for each node corresponds to a multiple logistic regression \citep{agresti2014categorical}:
\[
\mathcal{L}_i\left(\pmb{\tau}, \pmb{\Omega} ; \pmb{x} \right)   =  x_i \left(\tau_i  +  \sum_{j} \omega_{ij} x_j\right)  - \sum_{x_i} \exp \left( x_i \left(\tau_i  +  \sum_{j} \omega_{ij} x_j  \right) \right).
\]
Here, the subscript $i$ indicates that the likelihood function is based on the conditional probability for node $i$ given the other nodes. Instead of optimizing the full likelihood of \eqref{handbook:eq:IsingModel}, the pseudolikelihood (PL; \citealt{besag1975statistical}) can be optimized instead. The pseudolikelihood approximates the likelihood with the product of univariate conditional likelihoods in \eqref{handbook:eq:conditional}:
\[
\ln \mathrm{PL} = \sum_{i=1}^k \mathcal{L}_i\left(\pmb{\tau}, \pmb{\Omega} ; \pmb{x} \right)
\]
Finally, disjoint pseudolikelihood estimation can be used. In this approach, each conditional likelihood is optimized separately \citep{liu2012distributed}. This routine corresponds to repeatedly performing a multiple logistic regression in which one node is the response variable and all other nodes are the predictors; by predicting $x_i$ from $\pmb{x}^{(-i)}$ estimates can be obtained for $\pmb{\omega}_i$ and $\tau_i$. After estimating a multiple logistic regression for each node on all remaining nodes, a single estimate is obtained for every $\tau_i$ and two estimates are obtained for every $\omega_{ij}$--the latter can be averaged to obtain an estimate of the relevant network parameter. Many statistical programs, such as the \emph{R} function \textVerb{glm}, can be used to perform logistic regressions. Estimation of the Ising model via log-linear modeling, maximal pseudolikelihood, and repeated multiple logistic regressions and have been implemented in the \textVerb{EstimateIsing} function in the \emph{R} package \emph{IsingSampler} \citep{IsingSampler}.

While the above-mentioned methods of estimating the Ising model are tractable, they all require a considerable amount of data to obtain reliable estimates. For example, in log-linear analysis, cells in the $2^P$ contingency table that are zero---which will occur often if $N < 2^P$---can cause parameter estimates to grow to $\infty$ \citep{agresti2014categorical}, and in logistic regression predictors with low variance (e.g., a very hard item) can substantively increase standard errors \citep{whittaker2009graphical}. To estimate the Ising model, $P$ thresholds and $P(P-1)/2$ network parameter have to be estimated, while in standard log linear approaches, rules of thumb suggest that the sample size needs to be three times higher than the number of parameters to obtain reliable estimates. In psychometrics, the number of data points is often far too limited for this requirement to hold. To estimate parameters of graphical models with limited amounts of observations, therefore, regularization methods have been proposed \citep{meinshausen2006high,friedman2008sparse}.

When regularization is applied, a penalized version of the (pseudo) likelihood is optimized. The most common regularization method is $\ell_1$ regularization--commonly known as the least absolute shrinkage and selection operator (LASSO; \citealt{tibshirani1996regression})--in which the sum of absolute parameter values is penalized to be under some value. \citet{ravikumar2010high} employed $\ell_1$-regularized logistic regression to estimate the structure of the Ising model via disjoint maximum pseudolikelihood estimation. For each node $i$ the following expression is maximized \citep{Friedman2009}:
\begin{align}
\label{handbook:logpen}
\max_{\tau_i, \pmb{\omega}_i} \left[ \mathcal{L}_i\left(\pmb{\tau}, \pmb{\Omega} ; \pmb{x} \right) - \lambda \mathrm{Pen}\left( \pmb{\omega}_i \right) \right]
\end{align}
Where  $ \pmb{\omega}_i$ is the $i$th row (or column due to symmetry) of $\pmb{\Omega}$ and $\mathrm{Pen}\left( \pmb{\omega}_i \right)$ denotes the penalty function, which is defined in the LASSO as follows:
\[
\mathrm{Pen}_{\mathrm{\ell_1}}\left( \pmb{\omega}_i \right) = || \pmb{\omega}_i||_1 = \sum_{j=1, j!= i}^{k} |\omega_{ij}|
\]
The $\lambda$ in \eqref{handbook:logpen} is the regularization tuning parameter. The problem in above is equivalent to the constrained optimization problem:
\[
\max_{\tau_i, \pmb{\omega}_i} \left[ \mathcal{L}_i\left(\pmb{\tau}, \pmb{\Omega} ; \pmb{x} \right) \right], \quad \text{subject to }  || \pmb{\omega}_i||_1 < C
\]
in which $C$ is a constant that has a one-to-one monotone decreasing relationship with $\lambda$ \citep{lee2006efficient}. If $\lambda=0$, $C$ will equal the sum of absolute values of the maximum likelihood solution; increasing $\lambda$ will cause $C$ to be smaller, which forces the estimates of $\pmb{\omega}_i$ to shrink. Because the penalization uses absolute values, this causes parameter estimates to shrink to exactly zero. Thus, in moderately high values for $\lambda$ a sparse solution to the logistic regression problem is obtained in which many coefficients equal zero; the LASSO results in simple predictive models including only a few predictors.

\citet{ravikumar2010high} used LASSO to estimate the neighborhood---the connected nodes---of each node, resulting in an unweighted graph structure. In this approach, an edge is selected in the model if either $\omega_{ij}$ and $\omega_{ji}$ is nonzero (the OR-rule) or if both are nonzero (the AND-rule). To obtain estimates for the weights $\omega_{ij}$ and $\omega_{ji}$ can again be averaged. The $\lambda$ parameter is typically specified such that an optimal solution is obtained, which is commonly done through cross-validation or, more recently, by optimizing the extended Bayesian information criterion (EBIC; \citealt{chen2008EBIC, foygel2010extended, foygel2014high, van2014new}). 

In $K$-fold cross-validation, the data are subdivided in $K$ (usually $K=10$) blocks. For each of these blocks a model is fitted using only the remaining $K-1$ blocks of data, which are subsequently used to construct a prediction model for the block of interest. For a suitable range of $\lambda$ values, the predictive accuracy of this model can be computed, and subsequently the $\lambda$ under which the data were best predicted is chosen. If the sample size is relatively low, the predictive accuracy is typically much better for $\lambda>0$ than it is at the maximum likelihood solution of $\lambda=0$; it is preferred to regularize to avoid over-fitting. 

Alternatively, an information criterion can be used to directly penalize the likelihood for the number of parameters. The EBIC \citep{chen2008EBIC} augments the Bayesian information Criterion (BIC) with a hyperparameter $\gamma$ to additionally penalize the large space of possible models (networks):
\[
\mathrm{EBIC} = -2  \mathcal{L}_i\left(\pmb{\tau}, \pmb{\Omega} ; \pmb{x} \right) + \left| \pmb{\omega}_i \right| \ln\left(N\right) + 2 \gamma \left| \pmb{\omega}_i \right| \ln \left(k-1 \right)
\]
in which $ \left| \pmb{\omega}_i \right| $ is the number of nonzero parameters in $\pmb{\omega}_i$. Setting $\gamma = 0.25$ works well for the Ising model \citep{foygel2014high}. An optimal $\lambda$ can be chosen either for the entire Ising model, which improves parameter estimation, or for each node separately in disjoint pseudolkelihood estimation, which improves neighborhood selection. While $K$-fold cross-validation does not require the computation of the intractable likelihood function, EBIC does. Thus, when using EBIC estimation $\lambda$ need be chosen per node. We have implemented $\ell_1$-regularized disjoint pseudolikelihood estimation of the Ising model using EBIC to select a tuning parameter per node in the \emph{R} package \emph{IsingFit} \citep{IsingFit, van2014new}, which uses \emph{glmnet} for optimization \citep{Friedman2009}.

The LASSO works well in estimating sparse network structures for the Ising model and can be used in combination with cross-validation or an information criterion to arrive at an interpretable model. However, it does so under the assumption that the true model in the population is sparse. So what if reality is not sparse, and we would not expect many missing edges in the network? As discussed earlier in this chapter, the absence of edges indicate conditional independence between nodes; if all nodes are caused by an unobserved cause we would not expect missing edges in the network but rather a low-rank network structure. In such cases, $\ell_2$ regularization---also called ridge regression---can be used which uses a quadratic penalty function:
\[
\mathrm{Pen}_{\mathrm{\ell_2}}\left( \pmb{\omega}_i \right) = || \pmb{\omega}_i||_2 = \sum_{j=1, j!= i}^{k} \omega_{ij}^2
\]
With this penalty parameters will not shrink to exactly zero but more or less smooth out; when two predictors are highly correlated the LASSO might pick only one where ridge regression will average out the effect of both predictors. \citet{zou2005regularization} proposed a compromise between both penalty functions in the \emph{elastic net}, which uses another tuning parameter, $\alpha$, to mix between $\ell_1$ and $\ell_2$ regularization:
\[
\mathrm{Pen}_{\mathrm{\mathrm{Elastic Net}}}\left( \pmb{\omega}_i \right)  = \sum_{j=1, j!= i}^{k} \frac{1}{2} (1-\alpha) \omega_{ij}^2 + \alpha |\omega_{ij}|
\]
If $\alpha=1$, the elastic net reduces to the LASSO penalty, and if $\alpha=0$ the elastic net reduces to the ridge penalty. When $\alpha>0$ exact zeroes can still be obtained in the solution, and sparsity increases both with $\lambda$ and $\alpha$. Since moving towards $\ell_2$ regularization reduces sparsity, selection of the tuning parameters using EBIC is less suited in the elastic net. Crossvalidation, however, is still capable of sketching the predictive accuracy for different values of both $\alpha$ and $\lambda$. Again, the \emph{R} package \emph{glmnet} \citep{Friedman2009} can be used for estimating parameters using the elastic net. We have implemented a procedure to compute the Ising model for a range of $\lambda$ and $\alpha$ values and obtain the predictive accuracy in the \emph{R} package \emph{elasticIsing} \citep{elasticIsing}.

One issue that is currently debated is inference of regularized parameters. Since the distribution of LASSO parameters is not well-behaved \citep{Buhlmann:2011,Buhlmann:2013},  \citet{Meinshausen:2009} developed the idea of using repeated sample splitting, where in the first sample the sparse set of variables are selected, followed by multiple comparison corrected $p$-values in the second sample. Another interesting idea is to remove the bias introduced by regularization, upon which `standard' procedures can be used \citep{Geer2013}. As a result the asymptotic distribution of the so-called de-sparsified LASSO parameters is normal with the true parameter as mean and efficient variance (i.e., achieves the Cram\'{e}r-Rao bound).. Standard techniques are then applied and even confidence intervals with good coverage are obtained. The limitations here are (i) the sparsity level, which has to be $\le \sqrt{n/\ln(P)}$, and (ii) the 'beta-min' assumption, which imposes a lower bound on the value of the smallest obtainable coefficient \citep{Buhlmann:2011}.

Finally, we can use the equivalence between MIRT and the Ising model to estimate a low-rank approximation of the Ising Model. MIRT software, such as the \emph{R} package \emph{mirt} \citep{mirt}, can be used for this purpose. More recently, \citet{marsman2015bayesian} have used the equivalence also presented in this chapter as a method for estimating low-rank Ising model using Full-data-information estimation. A good approximation of the Ising model can be obtained if the true Ising model is indeed low-rank, which can be checked by looking at the eigenvalue decomposition of the elastic Net approximation or by sequentially estimating the first eigenvectors through adding more latent factors in the MIRT analysis or estimating sequentially higher rank networks using the methodology of \citet{marsman2015bayesian}.

\subsection{Example Analysis}

\begin{figure}
    \subfloat[\label{1a}]{%
      \includegraphics[width=0.5\textwidth,page=1]{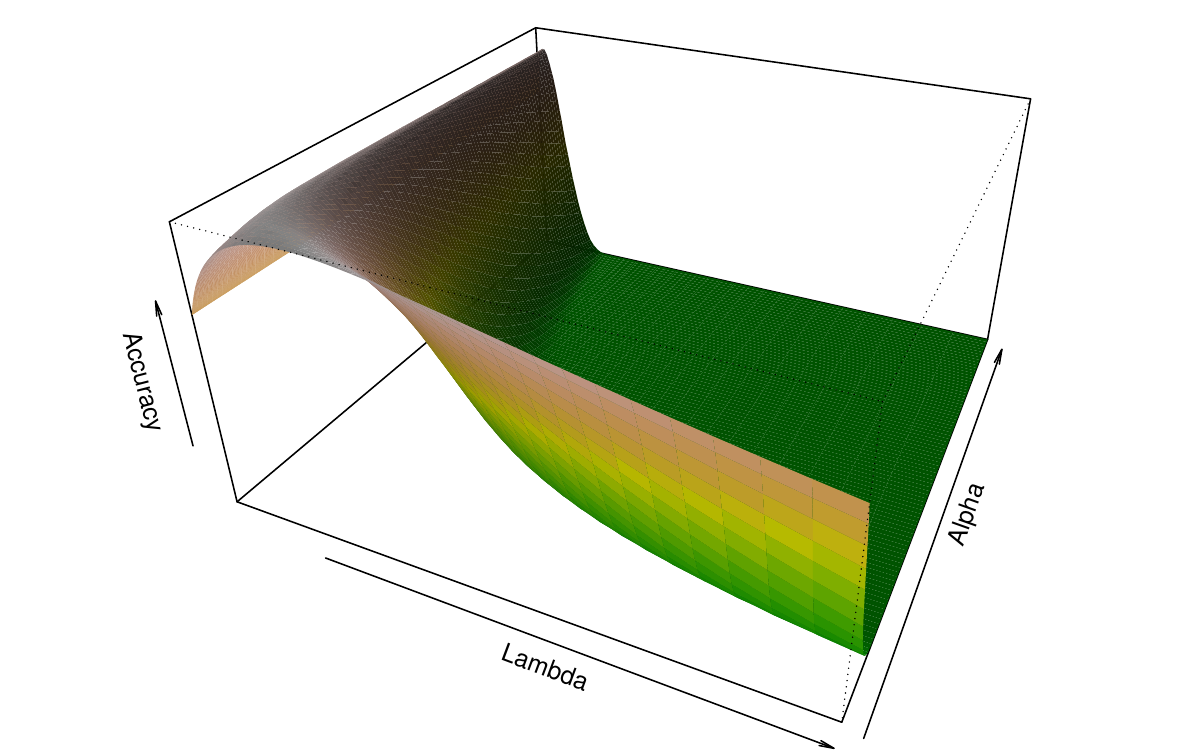}
    }
    \hfill
    \subfloat[\label{1b}]{%
      \includegraphics[width=0.5\textwidth,page=1]{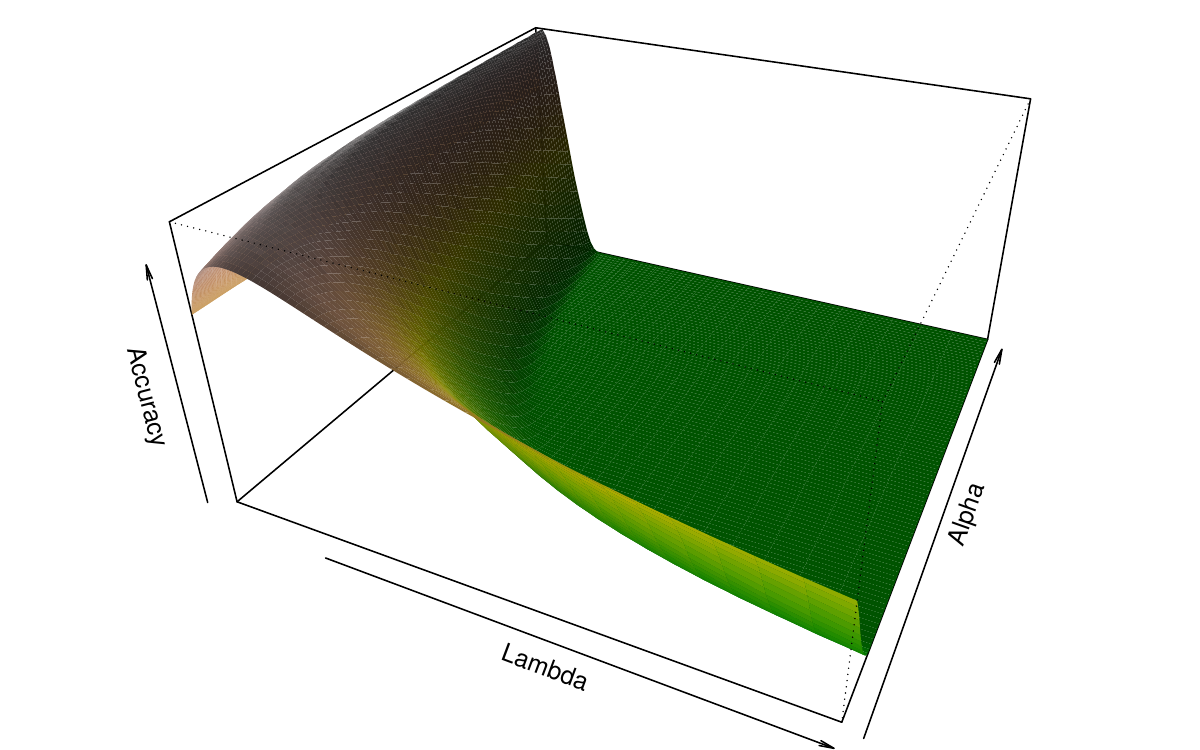}
    } \\
        \subfloat[\label{2a}]{%
      \includegraphics[width=0.5\textwidth,page=1]{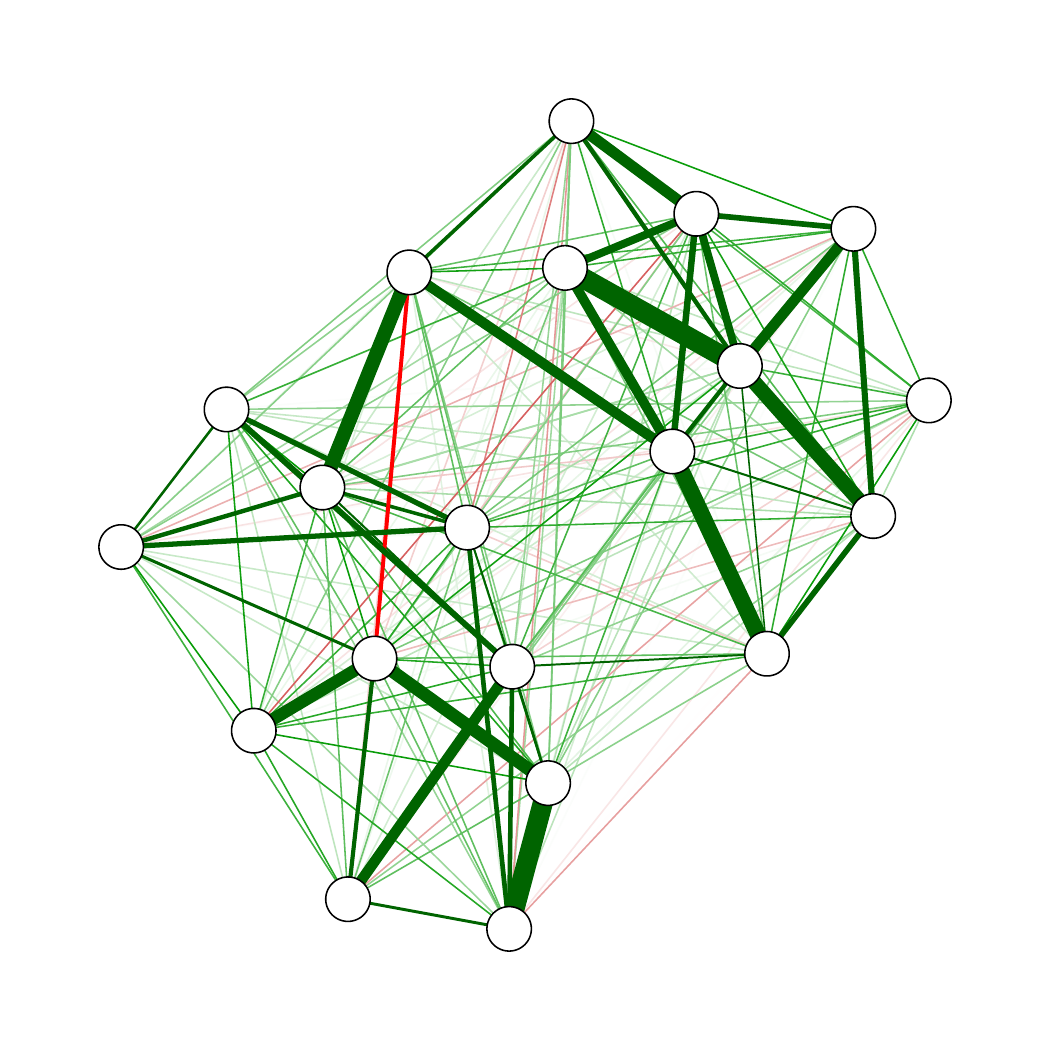}
    }
    \hfill
    \subfloat[\label{2b}]{%
      \includegraphics[width=0.5\textwidth,page=1]{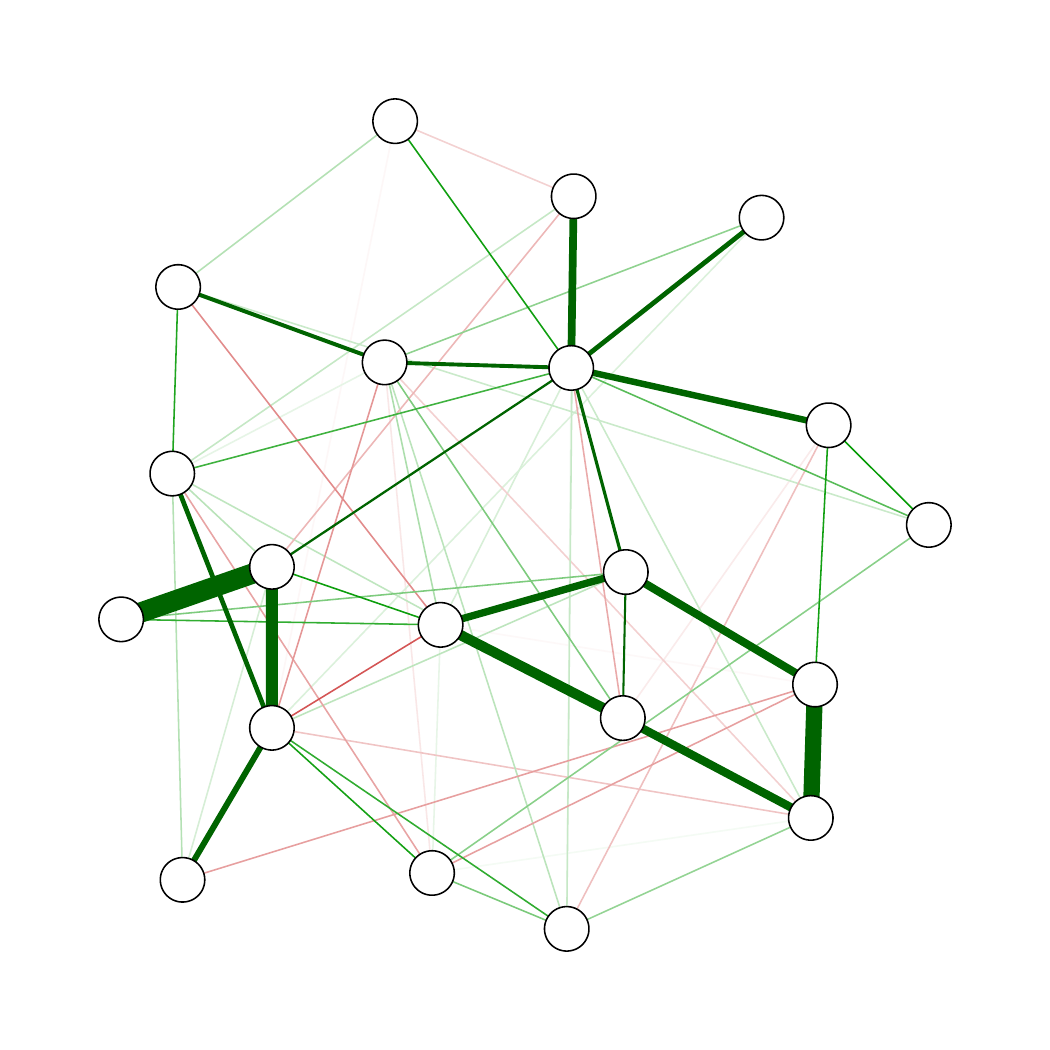}
    } \\
        \subfloat[\label{3a}]{%
      \includegraphics[width=0.5\textwidth,page=1]{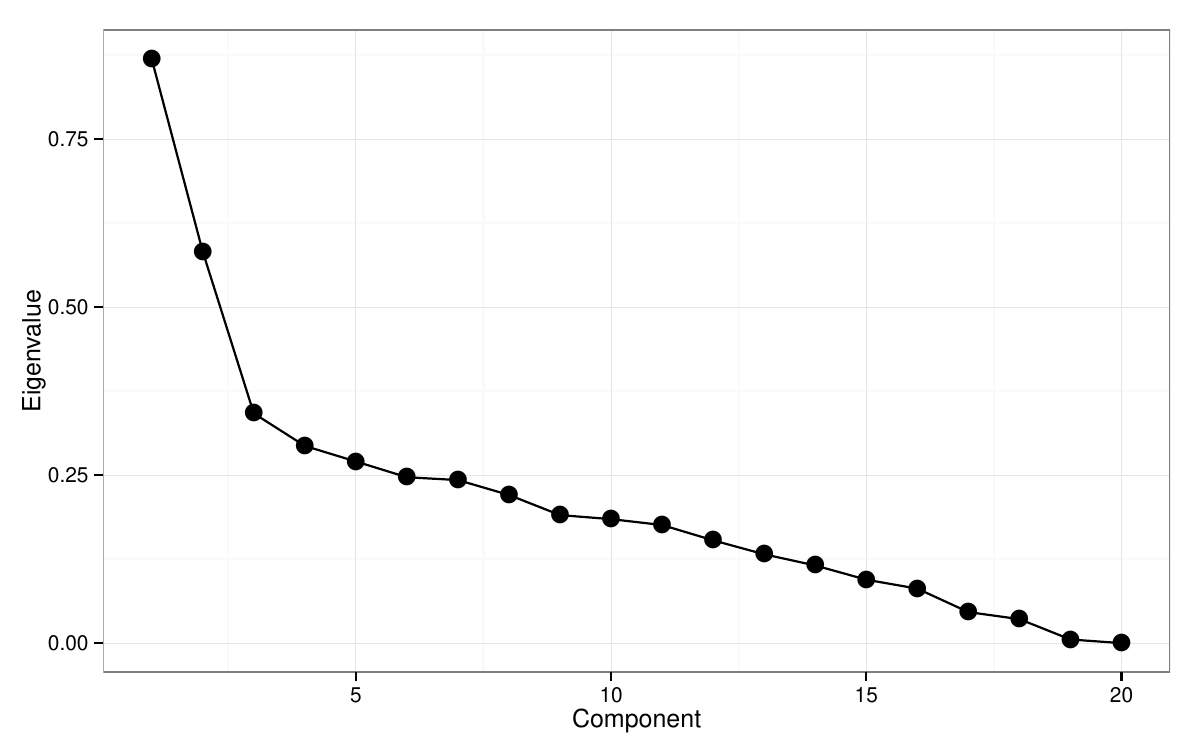}
    }
    \hfill
    \subfloat[\label{3b}]{%
      \includegraphics[width=0.5\textwidth,page=2]{fig5fg}
    } 
    \caption{Analysis results of two simulated datasets; left panels show results based on a dataset simulated according to a 2-factor MIRT Model, while right panels show results based on a dataset simulated with a sparse scale-free network. Panels (a) and (b) show the predictive accuracy under different elastic net tuning parameters $\lambda$ and $\alpha$, panels (c) and (d) the estimated optimal graph structures and panels (e) and (f) the eigenvalues of these graphs.}
    \label{handbook:fig:sim}
  \end{figure}
  
To illustrate the methods described in this chapter we simulated two datasets, both with $500$ measurements on $10$ dichotomous scored items. The first dataset, dataset A, was simulated according to a multidimensional Rasch model, in which the first five items are determined by the first factor and the last five items by the second factor. Factor levels where sampled from a multivariate normal distribution with unit variance and a correlation of $0.5$, while item difficulties where sampled from a standard normal distribution. The second dataset, dataset B, was sampled from a sparse network structure according to a Boltzmann Machine. A scale-free network was simulated using the Barabasi game algorithm \citep{barabasi1999emergence} in the \emph{R} package \emph{igraph} \citep{igraph} and a random connection probability of $5\%$. The edge weights where subsequently sampled from a uniform distribution between $0.75$ and $1$ (in line with the conception that most items in psychometrics relate positively with each other) and thresholds where sampled from a uniform distribution between $-3$ and $-1$. To simulate the responses the \emph{R} package \emph{IsingSampler} was used. The datasets where analyzed using the \emph{elasticIsing} package in \emph{R} \citep{elasticIsing}; $10$-fold cross-validation was used to estimate the predictive accuracy of tuning parameters $\lambda$ and $\alpha$ on a grid of $100$ logarithmically spaced $\lambda$ values between $0.001$ and $1$ and $100$ $\alpha$ values equally spaced between $0$ and $1$.  

Figure~\ref{handbook:fig:sim} shows the results of the analyses. The left panels show the results for dataset A and the right panel shows the result for dataset B. The top panels show the negative mean squared prediction error for different values of $\lambda$ and $\alpha$. In both datasets, regularized models perform better than unregularized models. The plateaus on the right of the graphs show the performance of the independence graph in which all network parameters are set to zero. Dataset A obtained a maximum accuracy at $\alpha=0$ and $\lambda=0.201$, thus in dataset A $\ell_2$-regularization is preferred over $\ell_1$ regularization, which is to be expected since the data were simulated under a model in which none of the edge weights should equal zero. In dataset B a maximum was obtained at $\alpha=0.960$ and $\lambda=0.017$, indicating that in dataset B regularization close to $\ell_1$ is preferred. The middle panels show visualizations of the obtained best performing networks made with the \emph{qgraph} package \citep{jssv048i04}; green edges represent positive weights, red edges negative weights and the wider and more saturated an edge the stronger the absolute weight. It can be seen that dataset A portrays two clusters while Dataset B portrays a sparse structure. Finally, the bottom panels show the eigenvalues of both graphs; Dataset A clearly indicates two dominant components whereas Dataset B does not indicate any dominant component. 

These results show that the estimation techniques perform adequately, as expected. As discussed earlier in this chapter, the eigenvalue decomposition directly corresponds to the number of latent variables present if the common cause model is true, as is the case in dataset A. Furthermore, if the common cause model is true the resulting graph should not be sparse but low rank, as is the case in the results on dataset A.

\section{The Interpretation of Latent Variables in Psychometric Models}

Since Spearman's \citeyearpar{spearman1904general} conception of general intelligence as the common determinant of observed differences in cognitive test scores, latent variables have played a central role in psychometric models. The theoretical status of the latent variable in psychometric models has been controversial and the topic of heated debates in various subfields of psychology, like those concerned with the study of intelligence (e.g., \citealt{jensen1998g}) and personality \citep{mccrae2008empirical}. The pivotal issue in these debates is whether latent variables posited in statistical models have referents outside of the model; that is, the central question is whether latent variables like $g$ in intelligence or ``extraversion'' in personality research refer to a property of individuals that exists independently of the model fitting exercise of the researcher \citep{borsboom2003theoretical, van2006dynamical, cramer2010comorbidity}. If they do have such independent existence, then the model formulation appears to dictate a causal relation between latent and observed variables, in which the former cause the latter; after all, the latent variable has all the formal properties of a common cause because it screens off the correlation between the item responses (a property denoted local independence in the psychometric literature; \citealt{borsboom2005measuring, reichenbach1991direction}). The condition of {\em vanishing tetrads}, that Spearman \citeyearpar{spearman1904general} introduced as a model test for the veracity of the common factor model is currently seen as one of the hallmark conditions of the common cause model \citep{bollen1991conventional}. 

This would suggest that the latent variable model is intimately intertwined with a so-called reflective measurement model interpretation \citep{edwards2000nature, howell2007reconsidering}, also known as an effect indicators model \citep{bollen1991conventional} in which the measured attribute is represented as the cause of the test scores. This conceptualization is in keeping with causal accounts of measurement and validity \citep{borsboom2003theoretical, markus2013reflective} and indeed seems to fit the intuition of researchers in fields where psychometric models dominate, like personality. For example, \citet{mccrae2008empirical} note that they assume that extraversion causes party-going behavior, and as such this trait determines the answer to the question ``do you often go to parties'' in a causal fashion. \citet{jensen1998g} offers similar ideas on the relation between intelligence and the {\em g}-factor. Also, in clinical psychology, Reise and Waller (\citeyear{reise2009item}, p. 26) note that ``to model item responses to a clinical instrument [with IRT], a researcher must first assume that the item covariation is caused by a continuous latent variable''. 

However, not all researchers are convinced that a causal interpretation of the relation between latent and observed variable makes sense. For instance, \citet{mcdonald2003behavior} notes that the interpretation is somewhat vacuous as long as no substantive theoretical of empirical identification of the latent variable can be given; a similar point is made by \citet{borsboom2013network}. That is, as long as the sole evidence for the existence of a latent variable lies in the structure of the data to which it is fitted, the latent variable appears to have a merely statistical meaning and to grant such a statistical entity substantive meaning appears to be tantamount to overinterpreting the model. Thus, the common cause interpretation of latent variables at best enjoys mixed support.

A second interpretation of latent variables that has been put forward in the literature is one in which latent variables do not figure as common causes of the item responses, but as so-called behavior domains. Behavior domains are sets of behaviors relevant to substantive concepts like intelligence, extraversion, or cognitive ability \citep{mulaik1978effect, mcdonald2003behavior}. For instance, one can think of the behavior domain of addition as being defined through the set of all test items of the form $x + y =\ldots$. The actual items in a test are considered to be a sample from that domain. A latent variable can then be conceptualized as a so-called {\em tail-measure} defined on the behavior domain \citep{ellis1997tail}. One can intuitively think of this as the total test score of a person on the infinite set of items included in the behavior domain. \citet{ellis1997tail} have shown that, if the item responses included in the domain satisfy the properties of monotonicity, positive association, and vanishing conditional independence, the latent variable can indeed be defined as a tail measure. The relation between the item responses and the latent variable is, in this case, not sensibly construed as causal, because the item responses are a part of the behavior domain; this violates the requirement, made in virtually all theories of causality, that cause and effect should be separate entities \citep{markus2013reflective}. Rather, the relation between item responses and latent variable is conceptualized as a sampling relation, which means the inference from indicators to latent variable is not a species of causal inference, but of statistical generalization.

Although in some contexts the behavior domain interpretation does seem plausible, it has several theoretical shortcomings of its own. Most importantly, the model interpretation appears to beg the important explanatory question of why we observe statistical associations between item responses. For instance, \citet{ellis1997tail} manifest conditions specify that the items included in a behavior domain should look exactly as if they were generated by a common cause; in essence, the only sets of items that would qualify as behavior domains are infinite sets of items that would fit a unidimensional IRT model perfectly. The question of why such sets would fit a unidimensional model is thus left open in this interpretation. A second problem is that the model specifies infinite behavior domains (measures on finite domains cannot be interpreted as latent variables because the axioms of  \citeauthor{ellis1997tail} will not be not satisfied in this case). In many applications, however, it is quite hard to come up with more than a few dozen of items before one starts repeating oneself (e.g., think of psychopathology symptoms or attitude items), and if one does come up with larger sets of items the unidimensionality requirement is typically violated. Even in applications that would seem to naturally suit the behavior domain interpretation, like the addition ability example given earlier, this is no trivial issue. Thus, the very property that buys the behavior domain interpretation its theoretical force (i.e., the construction of latent variables as tail measures on an infinite set of items that satisfies a unidimensional IRT model) is its substantive Achilles' heel.

Thus, the common cause interpretation of the latent variable model seems too make assumptions about the causal background of test scores that appear overly ambitious given the current scientific understanding of test scores. The behavior domain interpretation is much less demanding, but appears to be of limited use in situations where only a limited number of items is of interest and in addition offers no explanatory guidance with respect to answering the question why items hang together as they do. The network model may offer a way out of this theoretical conundrum because it specifies a third way of looking at latent variables, as explained in this chapter. As \citet{van2006dynamical} showed, data generated under a network model could explain the positive manifold often found in intelligence research which is often described as the $g$ factor or general intelligence; a $g$ factor emerged from a densely connected network even though it was not ``real''. This idea suggests the interpretation of latent variables as functions defined as cliques in a network of interacting components \citep{borsboom2011small,cramer2010comorbidity, cramer2012dimensions}. As we have shown in this chapter, this relation between networks and latent variables is quite general: given simple models of the interaction between variables, as encoded in the Ising model, one expects data that conform to psychometric models with latent variables. The theoretical importance of this result is that (a) it allows for a model interpretation that invokes no common cause of the item responses as in the reflective model interpretation, but (b) does not require assumptions about infinite behavior domains either. 

Thus, network approaches can offer a theoretical middle ground between causal and sampling interpretations of psychometric models. In a network, there clearly is nothing that corresponds to a causally effective latent variable, as posited in the reflective measurement model interpretation \citep{bollen1991conventional,edwards2000nature}. The network model thus evades the problematic assignment of causal force to latent variables like the g-factor and extraversion. These arise out of the network structure as epiphenomena; to treat them as causes of item responses involves an unjustified reification. On the other hand, however, the latent variable model as it arises out of a network structure does not require the antecedent identification of an infinite set of response behaviors as hypothesized to exist in behavior domain theory. Networks are typically finite structures that involve a limited number of nodes engaged in a limited number of interactions. Each clique in the network structure will generate one latent variable with entirely transparent theoretical properties and an analytically tractable distribution function. Of course, for a full interpretation of the Ising model analogous to that in physics, one has to be prepared to assume that the connections between nodes in the network signify actual interactions (i.e., they are not merely correlations); that is, connections between nodes are explicitly not spurious as they are in the reflective latent variable model, in which the causal effect of the latent variable produces the correlations between item responses. But if this assumption is granted, the theoretical status of the ensuing latent variable is transparent and may in many contexts be less problematic than the current conceptions in terms of reflective measurement models and behavior domains are.

Naturally, even though the Ising and IRT models have statistically equivalent representations, the interpretations of the model in terms of common causes and networks are not equivalent. That is, there is a substantial difference between the causal implications of a reflective latent variable model and of an Ising model. However, because for a given dataset the models are equivalent, distinguishing network models from common cause models requires the addition of (quasi-) experimental designs into the model. For example, suppose that in reality an Ising model holds for a set of variables; say we consider the depression symptoms ``insomnia'' and ``feelings of worthlessness''. The model implies that, if we were to causally intervene on the system by reducing or increasing insomnia, a change in feelings of worthlessness should ensue. In the latent variable model, in which the association between feelings of worthlessness and insomnia is entirely due to the common influence of a latent variable, an experimental intervention that changes insomnia will not be propagated through the system. In this case, the intervention variable will be associated only with insomnia, which means that the items will turn out to violate measurement invariance with respect to the intervention variable \citep{mellenbergh1989item, meredith1993measurement}. Thus, interventions on individual nodes in the system can propagate to other nodes in a network model, but not in a latent variable model. This is a testable implication in cases where one has experimental interventions that plausibly target a single node in the system. \citet{fried2013depression} have identified a number of factors in depression that appear to work in this way.

Note that a similar argument does not necessarily work with variables that are causal consequences of the observed variables. Both in a latent variable model and in a network model, individual observed variables might have distinct outgoing effects, i.e., affect unique sets of external variables. Thus, insomnia may directly cause bags under the eyes, while feelings of worthlessness do not, without violating assumptions of either model. In the network model, this is because the outgoing effects of nodes do not play a role in the network if they do not feed back into the nodes that form the network. In the reflective model, this is because the model only speaks on the question of where the systematic variance in indicator variables comes from (i.e., this is produced by a latent variable), but not on what that systematic variance causes. As an example, one may measure the temperature of water by either putting a thermometer into the water, or by testing whether one can boil an egg in it. Both the thermometer reading and the boiled egg are plausibly construed as effects of the temperature in the water (the common cause latent variable in the system). However, only the boiled egg has the outgoing effect of satisfying one's appetite.

In addition to experimental interventions on the elements of the system, a network model rather than a latent variable model allows one to deduce what would happen upon changing the connectivity of the system. In a reflective latent variable model, the associations between variables are a function of the effect of the latent variable and the amount of noise present in the individual variables. Thus, the only ways to change the correlation between items is by changing the effect of the latent variable (e.g., by restricting the variance in the latent variable so as to produce restriction of range effects in the observables) or by increasing noise in the observed variables (e.g., by increasing variability in the conditions under which the measurements are taken). Thus, in a standard reflective latent variable model, the connection between observed variables is purely a correlation, and one can only change it indirectly through the variable that have proper causal roles in the system (i.e., latent variables and error variables).

However, in a network model, the associations between observed variables are not spurious; they are real, causally potent pathways, and thus externally forced changes in connection strengths can be envisioned. Such changes will affect the behavior of the system in a way that can be predicted from the model structure. For example, it is well known that increasing the connectivity of an Ising model can change its behavior from being linear (in which the total number of active nodes grows proportionally to the strength of external perturbations of the system) to being highly nonlinear. Under a situation of high connectivity, an Ising network features tipping points: in this situation, very small perturbations can have catastrophic effects. To give an example, a weakly connected network of depression symptoms could only be made depressed by strong external effects (e.g., the death of a spouse), whereas a strongly connected network could tumble into a depression through small perturbations (e.g., an annoying phone call from one's mother in law). Such a vulnerable network will also feature very specific behavior; for instance, when the network is approaching a transition, it will send out early warning signals like increased autocorrelation in a time series \citep{scheffer2009early}. Recent investigations suggest that such signals are indeed present in time series of individuals close to a transition \citep{van2014critical}. Latent variable models have no such consequences.

Thus, there are at least three ways in which network models and reflective latent variable models can be distinguished: through experimental manipulations of individual nodes, through experimental manipulations of connections in the network, and through investigation of the behavior of systems under highly frequent measurements that allow one to study the dynamics of the system in time series. Of course, a final and direct refutation of the network model would occur if one could empirically identify a latent variable (e.g., if one could show that the latent variable in a model for depression items was in fact identical with a property of the system that could be independently identified; say, serotonin shortage in the brain). However, such identifications of abstract psychometric latent variables with empirically identifiable common causes do not appear forthcoming. Arguably, then, psychometrics may do better to bet on network explanations of association patterns between psychometric variables than to hope for the empirical identification of latent common causes.

\section{Conclusion}
The correspondence between the Ising model and the MIRT model offers novel interpretations of long standing psychometric models, but also opens a gateway through which the psychometric can be connected to the physics literature. Although we have only begun to explore the possibilities that this connection may offer, the results are surprising and, in our view, offer a fresh look on the problems and challenges of psychometrics. In the current chapter, we have illustrated how network models could be useful in the conceptualization of psychometric data. The bridge between network models and latent variables offers research opportunities that range from model estimation to the philosophical analysis of measurement in psychology, and may very well alter our view of the foundations on which psychometric models should be built. 

As we have shown, network models may yield probability distributions that are exactly equivalent to this of IRT models. This means that latent variables can receive a novel interpretation: in addition to an interpretation of latent variables as common causes of the item responses (Bollen \& Lennox, 1991; Edwards \& Bagozzi, 2000), or as behavior domains from which the responses are a sample \citep{ellis1997tail, mcdonald2003behavior}, we can now also conceive of latent variables as mathematical abstractions that are defined on cliques of variables in a network. The extension of psychometric work to network modeling fits current developments in substantive psychology, in which network models have often been motivated by critiques of the latent variable paradigm. This has for instance happened in the context of intelligence research \citep{van2006dynamical}, clinical psychology \citep{ cramer2010comorbidity, borsboom2013network}, and personality \citep{cramer2012dimensions,costantini2015state}. It should be noted that, in view of the equivalence between latent variable models and network models proven here, even though these critiques may impinge on the common cause interpretation of latent variable models, they do not directly apply to latent variable models themselves. Latent variable models may in fact fit psychometric data well {\em because} these data result from a network of interacting components. In such a case, the latent variable should be thought of as a convenient fiction, but the latent variable model may nevertheless be useful; for instance, as we have argued in the current chapter, the MIRT model can be profitably used to estimate the parameters of a (low rank) network. Of course, the reverse holds as well: certain network structures may fit the data because cliques of connected network components result from unobserved common causes in the data. An important question is under which circumstances the equivalence between the MIRT model and the Ising model breaks down, i.e., which experimental manipulations or extended datasets could be used to decide between a common cause versus a network interpretation of the data. In the current paper, we have offered some suggestions for further work in this direction, which we think offers considerable opportunities for psychometric progress.

As psychometrics starts to deal with network models, we think the Ising model offers a canonical form for network psychometrics, because it deals with binary data and is equivalent to well-known models from IRT. The Ising model has several intuitive interpretations: as a model for interacting components, as an association model with at most pairwise interactions, and as the joint distribution of response and predictor variables in a logistic regression. Especially the analogy between networks of psychometric variables (e.g., psychopathology symptoms such as depressed mood, fatigue, and concentration loss) and networks of interacting particles (e.g., as in the magnetization examples) offers suggestive possibilities for the construction of novel theoretical accounts of the relation between constructs (e.g., depression) and observables as modeled in psychometrics (e.g., symptomatology). In the current chapter, we only focused on the Ising model for binary data, but of course the work we have ignited here invites extensions in various other directions. For example, for polymotous data, the generalized Potts model could be used, although it should be noted that this model does require the response options to be discrete values that are shared over all variables, which may not suit typical psychometric applications. Another popular type of PMRF is the Gaussian Random Field (GRF; \citealt{lauritzen1996graphical}), which has exactly the same form as the model in \eqref{handbook:eq:matising} except that now $\boldsymbol{x}$ is continuous and assumed to follow a multivariate Gaussian density. This model is considerably appealing as it has a tractable normalizing constant rather than the intractable partition function of the Ising model. The inverse of the covariance matrix---the precision matrix---can be standardized as a partial correlation matrix and directly corresponds to the $\boldsymbol{\Omega}$ matrix of the Ising model. Furthermore, where the Ising model reduces to a series of logistic regressions for each node, the GRF reduces to a multiple linear regression for each node. It can easily be proven that also in the GRF the rank of the (partial) correlation matrix---cliques in the network---correspond to the latent dimensionality if the common cause model is true \citep{chandrasekaran2010latent}. A great body of literature exists on estimating and fitting GRFs even when the amount of observations is limited versus the amount of nodes \citep{meinshausen2006high, friedman2008sparse, foygel2010extended}. Furthermore, promising methods are now available for the estimation of a GRF even in non-Gaussian data, provided the data are continuous \citep{liu2009nonparanormal, liu2012high}.

\bibliographystyle{apalike}
\bibliography{Bibliography}

\appendix

\section{Proof of Equivalence Between the Ising Model and MIRT}

\label{handbook:proof}

To prove the equivalence between the Ising model and MIRT, we first need to rewrite the Ising Model in matrix form:
\begin{equation}
\label{handbook:eq:matising}
p(\pmb{X} = \pmb{x}) =  \frac{1}{Z} \exp\left( \pmb{\tau}^\top \pmb{x} + \frac{1}{2} \pmb{x}^\top\pmb{\Omega}\pmb{x} \right),
\end{equation}
in which $\pmb{\Omega}$ is an $P \times P$ matrix containing network parameters $\omega_{ij}$ as its elements, which corresponds in graph theory to the adjacency or weights matrix. Note that, in this representation, the diagonal values of $\pmb{\Omega}$ are used. However, since $x_i$ can be only $-1$ or $1$, $x_i^2 = 1$ for any combination, and the diagonal values are cancelled out in the normalizing constant $Z$. Thus, arbitrary values can be used in the diagonal of $\pmb{\Omega}$. Since $\pmb{\Omega}$ is a real and symmetrical matrix, we can take the usual eigenvalue decomposition:
\[
\pmb{\Omega} = \pmb{Q} \pmb{\Lambda} \pmb{Q}^\top,
\]
in which $\pmb{\Lambda}$ is a diagonal matrix containing eigenvalues $\lambda_1, \lambda_2, \ldots, \lambda_P$ on its diagonal, and $ \pmb{Q}$ is an orthonormal matrix containing eigenvectors $\pmb{q}_1, \ldots, \pmb{q}_P$ as its columns. Inserting the eigenvalue decomposition into \eqref{handbook:eq:matising} gives:
\begin{align}
\label{handbook:isingEigen}
p(\pmb{X} = \pmb{x}) &= \frac{1}{Z} \exp\left(\sum_i \tau_i x_i \right)  \prod_j \exp \left( \frac{\lambda_j}{2} \left(\sum_i q_{ij} x_i\right)^2 \right).
\end{align}
Due to the unidentified and arbitrary diagonal of $\pmb{\Omega}$ we can force $\pmb{\Omega}$ to be positive semi-definite---requiring all eigenvalues to be nonnegative---by shifting the eigenvalues with some constant $c$:
\[
\pmb{\Omega} + c\pmb{I} = \pmb{Q} \left( \pmb{\Lambda}  + c\pmb{I} \right)\pmb{Q}^\top.
\]

Following the work of \citet{kac1966mathematical}, we can use the following identity:
\begin{align*}
e^{y^2} &= \int_{-\infty}^{\infty} \frac{e^{-2ct - t^2}}{\sqrt{\pi}}  \, \mathrm{d}t  ,
\end{align*}
with $y = \sqrt{\frac{\lambda_j}{2}\left(\sum_i q_{ij} x_i\right)^2}$ and $t=\theta_j$ to rewrite \eqref{handbook:isingEigen} as follows:
{\footnotesize
\begin{align*}
p(\pmb{X} = \pmb{x})
&= \frac{1}{Z } \int_{-\infty}^{\infty} \frac{ \exp\left(\sum_j  - \theta_j^2\right)}{\sqrt{\pi^P}} \prod_i \exp\left( x_i \left( \tau_i  +  \sum_j -2\sqrt{\frac{\lambda_j}{2}} q_{ij} \theta_j \right) \right) \, \mathrm{d}\pmb{\theta}  .
\end{align*}
}
Reparameterizing $\tau_i = -\delta_i$ and $-2\sqrt{\frac{\lambda_j}{2}}q_{ij} = \alpha_{ij}$ we obtain:
\begin{align}
\label{handbook:eq:pIRTpar}
  p(\pmb{X} = \pmb{x}) &=   \int_{-\infty}^{\infty} \frac{1}{Z} \frac{ \exp\left(\sum_j  - \theta_j^2\right)}{ \sqrt{\pi^P}} \prod_i \exp\left(  x_i \left( \pmb{\alpha}_i^\top \pmb{\theta} - \delta_i  \right) \right) \, \mathrm{d}\pmb{\theta}  .
\end{align}
The same transformations can be used to obtain a different expression for $Z$:
\begin{align}
\nonumber
  Z
   &=  \int_{-\infty}^{\infty} \frac{ \exp\left(\sum_j  - \theta_j^2\right)}{ \sqrt{\pi^P}} \sum_{\pmb{x}} \prod_i \exp\left(  x_i \left( \pmb{\alpha}_i^\top \pmb{\theta} - \delta_i  \right) \right) \, \mathrm{d}\pmb{\theta}  \\
   \label{handbook:eq:Z}
   &= \int_{-\infty}^{\infty} \frac{ \exp\left(\sum_j  - \theta_j^2\right)}{ \sqrt{\pi^P}}  \prod_i \sum_{x_i}  \exp\left(  x_i \left( \pmb{\alpha}_i^\top \pmb{\theta} - \delta_i  \right) \right)  \, \mathrm{d}\pmb{\theta}  .
\end{align}
Finally, inserting \eqref{handbook:eq:Z} into \eqref{handbook:eq:pIRTpar}, multiplying by $\frac{\prod_i \sum_{x_i}  \exp\left(  x_i \left( \pmb{\alpha}_i^\top \pmb{\theta} - \delta_i  \right) \right) }{\prod_i \sum_{x_i}  \exp\left(  x_i \left( \pmb{\alpha}_i^\top \pmb{\theta} - \delta_i  \right) \right) } $, and rearranging gives:
\begin{align}
\nonumber
  p(\pmb{X} = \pmb{x}) &=  \int_{-\infty}^{\infty} \frac{ \frac{ \exp\left(\sum_j  - \theta_j^2\right) }{ \sqrt{\pi^{P}}  }  \prod_i \sum_{x_i}  \exp\left(  x_i \left( \pmb{\alpha}_i^\top \pmb{\theta} - \delta_i  \right) \right)  }{ \int_{-\infty}^{\infty}  \frac{\exp\left(\sum_j  - \theta_j^2\right) }{ \sqrt{\pi^{P}}} \prod_i \sum_{x_i}  \exp\left(  x_i \left( \pmb{\alpha}_i^\top \pmb{\theta} - \delta_i  \right) \right) \, \mathrm{d}\pmb{\theta} } \\
  \label{handbook:VolledigeMIRT}
  & \quad \cdot \prod_i \frac{   \exp\left(  x_i \left( \pmb{\alpha}_i^\top \pmb{\theta} - \delta_i  \right) \right) }{  \sum_{x_i}  \exp\left(  x_i \left( \pmb{\alpha}_i^\top \pmb{\theta} - \delta_i  \right) \right)  }   \, \mathrm{d}\pmb{\theta} .
\end{align}

The first part of the integral on the right hand side of \eqref{handbook:VolledigeMIRT} corresponds to a distribution that sums to $1$ for a $P$-dimensional random vector $\pmb{\Theta}$:
\[
f(\pmb{\theta}) \propto   \frac{ \exp\left(\sum_j  - \theta_j^2\right)}{ \sqrt{\pi^P}}    \prod_i \sum_{x_i}  \exp\left(  x_i \left( \pmb{\alpha}_i^\top \pmb{\theta} - \delta_i  \right) \right) ,
\]
and the second part corresponds to the 2-parameter logistic MIRT probability of the response vector as in \eqref{handbook:MIRTprob}:
\[
P( \pmb{X} = \pmb{x} \mid \pmb{\Theta} = \pmb{\theta}) = \prod_i \frac{   \exp\left(  x_i \left( \pmb{\alpha}_i^\top \pmb{\theta} - \delta_i  \right) \right) }{ \sum_{x_i}  \exp\left(  x_i \left( \pmb{\alpha}_i^\top \pmb{\theta} - \delta_i  \right) \right)  }.
\]
We can look further at this distribution by using Bayes' rule to examine the conditional distribution of $\pmb{\theta}$ given $\pmb{X} = \pmb{x}$:
\begin{align*}
f(\pmb{\theta} \mid \pmb{X} = \pmb{x}) &\propto   \Pr \left( \pmb{X} = \pmb{x} \mid \pmb{\Theta} = \pmb{\theta}  \right) f\left( \pmb{\theta}  \right) \\
&\propto     \exp\left( \pmb{x}^{\top}\pmb{A}\pmb{\theta}   - \pmb{ \theta}^{\top}  \pmb{ \theta}\right) \\
&\propto     \exp\left(    - \frac{1}{2} \left(\pmb{\theta} -  \frac{1}{2}\pmb{A}^{\top} \pmb{x}  \right)^{\top} 2\pmb{I} \left(\pmb{\theta} -  \frac{1}{2}\pmb{A}^{\top} \pmb{x}  \right) \right)\\
\end{align*}
and see that the posterior distribution of $\pmb{\Theta}$ is a multivariate Gaussian distribution:
\begin{equation}
\label{handbook:eq:post}
\pmb{\Theta} \mid \pmb{X} =  \pmb{x} \sim N_P \left( \pm \frac{1}{2} \boldsymbol{A}^\top \boldsymbol{x}, \sqrt{\frac{1}{2}}\pmb{I}  \right),
\end{equation}
in which $\pmb{A}$ is a matrix containing the discrimination parameters $\pmb{\alpha}_i$ as its rows and $\pm$ indicates that columns $\boldsymbol{a}_j$ could be multiplied with $-1$ due to that both the positive and negative root can be used in$\sqrt{\frac{\lambda_j}{2}}$, simply indicating whether the items overall are positively or negatively influenced by the latent trait $\boldsymbol{\theta}$. Additionally, Since the variance--covariance matrix of $\boldsymbol{\theta}$ equals zero in all nondiagonal elements, $\boldsymbol{\theta}$ is orthogonal. Thus, the multivariate density can be decomposed as the product of univariate densities:
\[
\Theta_j \mid \boldsymbol{X} = \boldsymbol{x} \sim N\left( \pm \frac{1}{2} \sum_i a_{ij} x_i ,   \sqrt{\frac{1}{2}} \right).
\]

\section{Glossary of Notation}
\label{handbook:glossary}

{\footnotesize
\begin{center}
\begin{tabular}{c L{3cm} L{5cm}}
Symbol & Dimension & Description \\
\hline
$ \left\{ \ldots \right\}$ & & Set of distinct values. \\
$ \left(a, b  \right)$ & & Interval between $a$ and $b$. \\
$P$ & $\mathbb{N}$  & Number of variables. \\
$N$ & $\mathbb{N}$  & Number of observations. \\
$\boldsymbol{X}$ & $\left\{-1, 1 \right\}^P$  & Random vector of binary variables. \\ 
$\boldsymbol{x}$ & $\left\{-1, 1 \right\}^P$  & A possible realization of $\boldsymbol{X}$. \\ 
$n( \pmb{x} )$ & $\mathbb{N}$  & Number of observations with response pattern $\pmb{x}$. \\
$i$, $j$, $k$ and $l$ &  $\left\{ 1,2,\ldots,P\right\}, j \not= i$ & Subscripts of  random variables. \\
$\boldsymbol{X}^{-(i)}$ & $\left\{-1, 1 \right\}^{P-1}$ & Random vector of binary variables without $X_i$. \\
$\boldsymbol{x}^{-(i)}$ & $\left\{-1, 1 \right\}^{P-1}$&  A possible realization of $\boldsymbol{X}^{-(i)}$. \\
$\boldsymbol{X}^{-(i,j)}$ & $\left\{-1, 1 \right\}^{P-2}$ & Random vector of binary variables without $X_i$ and $X_j$. \\
$\boldsymbol{x}^{-(i,j)}$ & $\left\{-1, 1 \right\}^{P-2}$&  A possible realization of $\boldsymbol{X}^{-(i)}$. \\
$\Pr\left( \ldots \right)$ & $\to \left(0, 1 \right)$ & Probability function. \\
$\phi_i(x_i)$ &  $\left\{-1, 1\right\} \to \mathbb{R}_{>0}$ & Node potential function. \\
$\phi_i(x_i, x_j)$ & $\left\{-1, 1\right\}^2 \to \mathbb{R}_{>0}$ & Pairwise potential function. \\
$\tau_i$ & $\mathbb{R}$ & Threshold parameter for node $X_i$ in the Ising model. Defined as $\tau_i = \ln \phi_i(1)$. \\
$\pmb{\tau}$ & $\mathbb{R}^P$ & Vector of threshold parameters, containing $\tau_i$ as its $i$th element. \\
$\omega_{ij}$ & $\mathbb{R}$ & Network parameter between nodes $X_i$ and $X_j$ in the Ising model. Defined as $\omega_{ij} = \ln \phi_{ij}(1,1)$. \\
$\pmb{\Omega}$ & $\mathbb{R}^{P \times P}$ and symmetrical & Matrix of network parameters, containing $\omega_{ij}$ as its $ij$th element. \\
$\pmb{\omega}_i$ & $\mathbb{R}^{P}$ & The $i$th row or column of $\pmb{\Omega}$. \\
$\mathrm{Pen}\left(\pmb{\omega}_i\right)$ & $\mathbb{R}^{P} \to \mathbb{R}$ & Penalization function of $\pmb{\omega}_i$. \\
$ \beta $ & $\mathbb{R}_{>0}$ & Inverse temperature in the Ising model. \\
$ H(\pmb{x})$ & $\left\{-1, 1 \right\}^P \to \mathbb{R}$ & Hamiltonian function denoting the energy of state $\pmb{x}$ in the Ising model. \\
$\nu_{\ldots}(\ldots)$ & $\to \mathbb{R}$ & The log potential functions, used in loglinear analysis. \\
$M$ & $\mathbb{N}$ & The number of latent factors. \\
$\boldsymbol{\Theta}$ & $\mathbb{R}^M$ & Random vector of continuous latent variables. \\ 
$\boldsymbol{\theta}$ & $\mathbb{R}^M$ & Realization of $\pmb{\Theta}$. \\ 
$\mathcal{L}\left(\pmb{\tau}, \pmb{\Omega} ; \pmb{x} \right)$ & $\to \mathbb{R}$  & Likelihood function based on $\Pr\left(\pmb{X} = \pmb{x} \right)$. \\
$\mathcal{L}_i\left(\pmb{\tau}, \pmb{\Omega} ; \pmb{x} \right)$ & $\to \mathbb{R}$  & Likelihood function based on $\Pr\left(X_i = x_i \mid \pmb{X}^{-(i)} = \pmb{x}^{-(i)} \right)$. \\
$\lambda$ & $\mathbb{R}_{>0}$ & LASSO tuning parameter \\
$\alpha$ & $(0,1)$ & Elastic net tuning parameter \\
\end{tabular}
\end{center}
}

\end{document}